\documentclass[aps,twocolumn,prb,amsmath,amssymb]{revtex4-1}

\usepackage{soul,color}

\usepackage{verbatim}
\usepackage[varg]{txfonts}
\usepackage{color}
\usepackage[colorlinks=true, citecolor=blue]{hyperref}
\usepackage{amsmath}
\usepackage{graphicx}
\usepackage{bm}
\usepackage{tikz}
\usetikzlibrary{arrows}
\usetikzlibrary{shapes}

\usepackage{dsfont}
\usepackage{calc}

\makeatletter
\def\textSq#1{%
	\begingroup
	\setlength{\fboxsep}{0.3ex}
	\setbox1=\hbox{#1}
	\setlength{\@tempdima}{\maxof{\wd1}{\ht1+\dp1}}
	\setlength{\@tempdimb}{(\@tempdima-\ht1+\dp1)/2}
	\raise-\@tempdimb\hbox{\fbox{\vbox to \@tempdima{%
				\vfil\hbox to \@tempdima{\hfil\copy1\hfil}\vfil}}}%
	\endgroup%
}
\makeatother

\def\XXint#1#2#3{{\setbox0=\hbox{$#1{#2#3}{\int}$}
    \vcenter{\hbox{$#2#3$}}\kern-.5\wd0}}

\def\be{\begin{equation}}
\def\ee{\end{equation}}

\def\bi{\begin{itemize}}
    \def\ei{\end{itemize}}
\def\bn{\begin{enumerate}}
    \def\en{\end{enumerate}}
\def\bea{\begin{eqnarray}}
\def\eea{\end{eqnarray}}
\newcommand{\bpm}{\begin{pmatrix}}
    \newcommand{\epm}{\end{pmatrix}}

\def\ba{\begin{array}}
    \def\ea{\end{array}}
\def\bd{\begin{displaymath}}
\def\ed{\end{displaymath}}

\renewcommand{\imath}{\hspace{1pt}\mathrm{i}\hspace{1pt}}

\renewcommand{\Im}{\mathop{\mathrm{Im}}\nolimits} 

\begin{document}

\title{The Kekulé-Kitaev model: linear and non-linear responses and magnetic field effects}


\author{Mohammad-Kazem Negahdari}
\affiliation{Department of Physics, Sharif University of Technology, Tehran 14588-89694, Iran}

\author{Abdollah Langari}
\email[]{langari@sharif.edu}
\affiliation{Department of Physics, Sharif University of Technology, Tehran 14588-89694, Iran}

\begin{abstract}

The Kekulé-Kitaev model, an extension of the Kitaev model, exhibits quantum spin liquid (QSL) properties, which  has an exact solution through Kitaev parton construction. In this study, we calculate the dynamical spin structure factor as a linear response and the third-order magnetic susceptibility as a nonlinear response using two-dimensional coherent spectroscopy for Kekulé-Kitaev model. Our results reveal that the few-matter fermion excitations approximation provides reliable results  for both linear and nonlinear responses. Notably, while the Kekulé-Kitaev model shows linear and nonlinear responses qualitatively similar to the Kitaev model, it displays distinct behavior under a weak uniform/staggered magnetic field.  Specifically, the Kekulé-Kitaev model does not present a non-Abelian phase under a uniform magnetic field, while such phase appears in the presence of a staggered magnetic field. Interestingly, within the non-Abelian phase, signals originating from two non-adjacent fluxes in the nonlinear response are stronger than signals from other flux excitations.
Furthermore, We demonstrate that the ground state of the Kekulé-Kitaev model at isotropic coupling is mapped, through unitary spin rotations, to an excited state of the Kitaev model with a uniform flux configuration.

\end{abstract}

\maketitle

\section{ Introduction}\label{Introduct}
In 1973, Anderson  introduced the idea  that quantum fluctuations could be sufficiently strong to stabilize a state in which long-range magnetic order fails to develop, even at absolute zero temperature\cite{anderson_1973}.  Since then, there have been ongoing efforts to understand and experimentally identify this phase of matter, which is now recognized as a quantum spin liquid (QSL). 
 However, the absence of long-range order is not sufficient to fully characterize the quantum spin liquid phase. Such exotic state is instead defined by the presence of fractionalized excitations and emergent gauge fields, which serve as key distinguishing features in the lack of conventional order
\cite{Mila_2000,Balents_2010,Knolle_field_2018, kivleson_QSL_2020,Lancaster2023}. The complexities involved in QSLs make relevance the use of gauge theories to understand their properties. QSLs are typically classified into categories such as $SU(2)$, $U(1)$, or $Z_2$, based on their underlying gauge structures \cite{Savary_2016,Zhou,Suchdev2023_book}. Geometric and exchange frustration are two factors that amplify the strong quantum fluctuations necessary for stabilizing QSLs. Organic compounds
$\kappa-(\text{ET})_2\text{Cu}_2(\text{CN})_3$\cite{Shimizu2003} and 
$\text{EtMe}_3\text{Sb}[\text{Pd}(\text{dimt})_2]_2$\cite{Itou2008}, with a triangular layered structure, and herbertsmithite 
$\text{ZnCu}_3\text{(OH)}_6\text{Cl}_2 $\cite{Shores2005}, with a kagome layered structure, are the most extensively studied candidates for quantum spin liquids with geometric frustration.

The Kitaev honeycomb lattice \cite{Kitaev_2006} illustrates the QSL phase arising from exchange frustration. This is an outstanding model,  for two reasons. First, it can be exactly solved by representing spins with four Majorana fermion operators. The resulting Hamiltonian describes itinerant Majorana fermions that freely move within a background of $Z_2$ gauge fields, defined by the localized Majorana fermions. Second, Jackeli and Khaliullin\cite{Jackeli_Khaliullin2009} introduced a mechanism that leads to the dominant Kitaev magnetic interaction in real materials. Based on the aforementioned mechanism, various candidates have been
\begin{figure}[!h]
	\includegraphics[width=0.95\columnwidth]{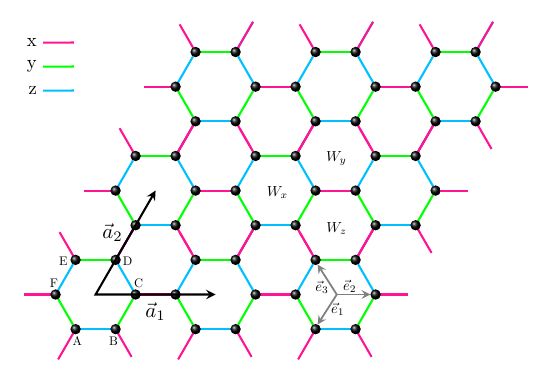}
	\caption{
		The Kekulé-Kitaev model. There are six sites in each unit cell labeled with the letters A, B, C, D, E, and F. There are three types of plaquette operators denoted by $W_x, W_y$ and $W_z$, which are constant of motions (see Eq.\ref{platelet_operator}). Different colors represent Ising interactions along $x$, $y$, and  $z$ direction.
		$\overrightarrow{\boldsymbol{a}}_1$ and $\overrightarrow{\boldsymbol{a}}_2$ are the lattice primitive vectors.
	}
	\label{lattice}
\end{figure}	
 proposed\cite{Jackeli_Khaliullin2,Winter2016_kitaevm,Trebst_kitaevm,Winter_2017_kitaevm,Takagi_Takayama2019,Liu_2021_3d_kitaevm,Gaoting_Lin2021,Sanders_Mole_2022,Yao_Zhao_2023,Abdeldaim2024}, with $\alpha$-$\text{RuCl}_3$ being one of the most significant Kitaev material\cite{Plumb_Clancy2014}. The representation of Majorana fermions extends the Hilbert space, which consists of two components: matter and gauge degrees of freedom. The matter excitations are described by itinerant Majorana fermions, while the gauge excitations, i.e. vortices or visons, are described by the localized Majorana fermions.

Different aspects of QSL have been studied to provide more information and clear identification of this elusive phase \cite{Knolle_field_2018,Lacroix2011_book,Knolle_Raman_Scattering_2014,Perreault_Brent_Knolle_2015,Halasz_Gabor_2016,Nasu2016,Savary_2017,Wellm2018,Kasahara_half,Bolens2018,Wen2019,Khatua2023}. In Neutron scattering \cite{Knolle_Chalker_Moessner2014,Knolle_Chalker_Moessner2015,Banerjee_Bridges_Nature_2016,Banerjee_Jiaqiang_Science_2017,Banerjee_Lampen_npj_2018}, which is a principal probe to study magnetic structures, the spectrum of QSL shows a broad and diffuse pattern due to spin fractionalization and the absence of long-range magnetic order.  The more we resolve this continuum of magnetic excitations, the closer we move towards an unambiguous and definitive identification of the spin liquid phase in experiment. Two-dimensional coherent spectroscopy (2DCS)\cite{Jonas_Review_2DCS,Cho_2DCS_2008,Hamm_book_2011,Woerner_2013,Cundiff_Mukamel2013,Lu_Li_2DCS_2017_spin_wave,Cho_book_2019}, which provides coherent interaction between light and matter, has demonstrated the ability to resolve the continuum of fractional excitations observed in the linear response. In fact, the inhomogeneous broadening caused by the multitude of fractional excitations is resolved by a signal known as the echo signal.
 Specifically,  itinerant and localized Majorana fermions are proposed to be observed using 2DCS for Kitaev model \cite{Armitage_nonlin,choi1}.  Sharp and informative nonlinear response provided by 2DCS, makes this probe an active and attractive direction for research \cite{Nandkishore_Choi2021,Li_Oshikawa_Lensing2021,Phuc_Trung_Direct2021,Fava_Biswas2021,marginal_Fermi,Oliver_Hart,Fava_Gopalakrishnan,Parameswaran_non1,Sim_GiBaik,Gao_Qi,Kanega_lin_nonlin,Negahdari_Langari_2023,Li_Zi_Yuan2023,Mootz_Orth2023,Watanabe_Trebst2023,Brenig2023,Qiang_orth2023,Emily2024,Watanabe_Trebst2024}.

In the Kitaev model, three types of bonds exist at each vertex of the honeycomb lattice. Its solvability in the language of Majorana fermions originates from this particular type of bond covering. However, with this feature, other bond coverings can also be considered\cite{Kamfor_2010}, in which each lattice vertex is the common point of three types of bonds. Here, we study the bond covering shown in Fig.~\ref{lattice}, known as the Kekulé-Kitaev model\cite{Kekule_Kitaev_mossner}. Away from the isotropic coupling, the Kekulé-Kitaev model transforms into the Kitaev toric code on the kagome lattice\cite{Kamfor_2010}. Moreover, it has two degenerate gapless Dirac cones in the center of the Brillouin zone (BZ) at the isotropic couplings. The inclusion of Heisenberg interactions results in a continuous quantum phase transition to a magnetically ordered phase, governed by a quantum order-by-disorder mechanism. This transition is expected to fall within the $3D$-$XY\times Z_2$ universality class.  Multi-band spectrum of the Kekulé-Kitaev model may provide a valuable framework for interpreting experimental data in spin-liquid candidates\cite{Mojarabian2020}.

Despite the existence of an exact solution for Kekulé-Kitaev model, less is known about the complexity of solution and its corresponding features on the exotic properties of the model. The extension of Hilbert space
within parton construction is classified to physical and unphysical states.
We unveil that contrary to the Kitaev model\cite{Zschocke_Vojta}, the physical ground state of the Kekulé-Kitaev model does not have odd parity for matter excitations in the trivial gauge configuration.
The calculation, which leads to 2DCS takes into account all of elementary excitations of the model and accordingly becomes sophisticated. However, for the Kitaev model 
the validity of the single-matter fermion excitations approximation \cite{Knolle_Chalker_Moessner2014,Knolle_Chalker_Moessner2015,Knolle2016_theses} dramatically simplifies the calculations of the physical responses. 
Nevertheless , the validity of single-matter fermion approximation for the Kekulé-Kitaev model has not been resolved.
To address the performance of single-matter fermion approximation
we calculate the contributions of higher order of excitations, i.e., the three-matter fermion excitations, in the dynamical spin structure factor. Our results justify that the higher-order contributions are negligible compared with the single-matter fermion ones. It paves the way to obtain the dynamical spin  structure factor and 2DCS of the Kekulé-Kitaev model within single-matter fermion approximation.

Moreover, we show that a weak uniform magnetic field in the $(1 1 1)$-direction does not open a gap in Kekulé-Kitaev model, contrary to the Kitaev model, where the magnetic field makes gapless Dirac cones to be massive. 
Hence, an interesting and relevant question is how one can open a gap in the spectrum of the Kekulé-Kitaev model at the isotropic couplings?
We find that exposing the system to a weak staggered magnetic field opens a gap in the spectrum. In this case, the gauge excitations (vortices) induce Majorana bound states inside the gap, which are responsible for the non-Abelian statistics of the vortices \cite{lahtinen2008,Lahtinen_Interacting_2011}. Therefore, we can induce a non-Abelian phase in the Kekulé-Kitaev model by applying a staggered magnetic field.

This paper is organized in following four sections: In Sec.~\ref{model}, we describe the diagonalization of the original Kekulé-Kitaev model in both real and momentum spaces and derive a relation for identifying physical and unphysical states within the Majorana representation. Furthermore, this section discusses the effects of weak uniform and staggered magnetic fields on the  Kekulé-Kitaev model and the stability of Dirac cones. In Secs.~\ref{linear_response_sec} and \ref{nonlinear_response_sec}, we present our results concerning the linear and nonlinear dynamical responses, respectively. Finally, Sec.~\ref{conclusion} is dedicated to summarizing our findings and presenting further discussions.

\section{Kekulé-Kitaev model and its exact solution}\label{model}
\begin{figure}[!t]
	\includegraphics[width=\columnwidth]{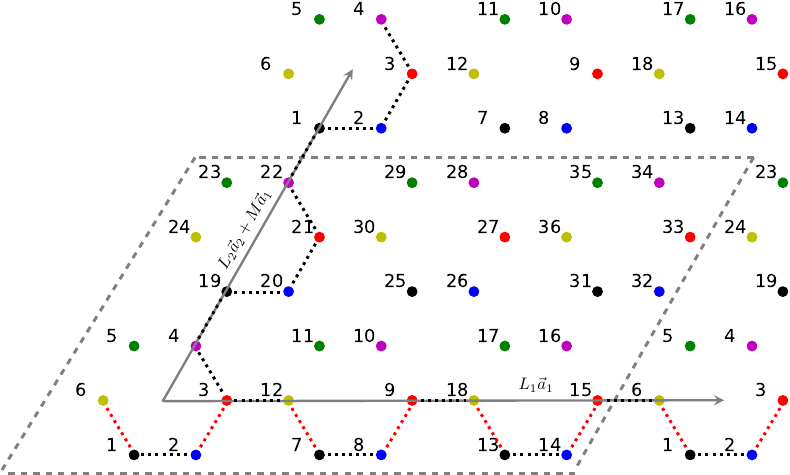}
	\caption{
		The Kekulé-Kitaev lattice with geometrical parameters $L_1=3$, $L_2=2$, and $M=0$. Dotted links show the topological operators in directions $\boldsymbol{a}_1$ and $\boldsymbol{a}_2$. Red  dotted links indicate the value of $u_{ij}=-1$. This is a gauge configuration for the $0$-flux sector with topological label $(+1,-1)$.
	}
	\label{topological_loop}
\end{figure}

The Kekulé-Kitaev model is a variant of the Kitaev model in which the distribution of Ising interaction on each hexagonal plaquette resembles the  Kekulé  pattern, as shown in Fig.~\ref{lattice}.  
The Hamiltonian is constructed with bond-dependent Ising interactions,
\begin{equation}
	\label{model_Ham}
	\hat{H}=-\sum_{\langle ij\rangle_{\alpha}}J_{\alpha}\hat{\sigma}^{\alpha}_i\hat{\sigma}^{\alpha}_j,
\end{equation}
where $ \hat{\sigma}_i^{\alpha} $ 
is the $\alpha$-component of Pauli matrix. This model has six sites per unit cell, labeled with letters A to F, as shown in Fig.~\ref{lattice}. For the Kekulé-Kitaev model unlike the Kitaev one, there exist three types of plaquettes: x/y/z-plaquette, where all outgoing bonds in the $\alpha $-plaquette are of $\alpha $-type . Like the Kitaev model, for each $\alpha$-plaquette, the flux operator 
 \begin{equation}
 	 \hat{W}_{\alpha}= \prod_{\substack{\text{link }\langle ij\rangle_{\gamma}\\\in\  \alpha\text{-plaquette}}}\hat{\sigma}^{\gamma}_i\hat{\sigma}^{\gamma}_j,
 	 \label{platelet_operator}
 \end{equation}
 is a constant of motion. We consider an x-plaquette as the unit cell. To impose an arbitrary periodic boundary condition, one can consider the model on a torus with the  basis vectors
 $ L_1\boldsymbol{a}_1 $ 
 and  
 $ L_2\boldsymbol{a}_2+M\boldsymbol{a}_1 $, see  [\onlinecite{Pedrocchi_Loss}] and [\onlinecite{Zschocke_Vojta}] for a similar approach and see appendix \ref{SiteNoPhase} for details  of the site labeling  convention. As an example Fig.~\ref{topological_loop} shows the case $ L_1=3, L_2=2 $, and $M=0$.  Apart from the lattice parameters $L_1$ and $L_2$, the parameter $M$ allows to define
any desired periodic boundary condition, which induces a twist in the lattice on the torus.

According to Kitaev's parton construction, each spin ($\sigma^{\alpha}$) is represented by three gauges 
$(\hat{b}_i^x,\hat{b}_i^y,\hat{b}_i^z) $  
and one matter ($ \hat{c}_i $) Majorana fermions,
\begin{equation}
	\label{partons}
	\hat{\sigma}_i^{\alpha}=i\hat{b}_i^{\alpha}\hat{c}_i,\quad\alpha=x,y,z.
\end{equation}
\noindent This representation  extends the Hilbert space of the Hamiltonian (\ref{model_Ham}), which is finally written in the following form,
\begin{equation}
	\label{u_ham}
	\tilde{H}_{\hat{u},\hat{c}}=\frac{i}{4}\sum_{\langle ij\rangle_{\alpha}}2J_{\alpha}\hat{u}_{\langle ij\rangle_{\alpha}}\hat{c}_i\hat{c}_j,
\end{equation}
where 
$ \hat{u}_{\langle ij\rangle_{\alpha}}=i\hat{b}^{\alpha}_i\hat{b}^{\alpha}_j $ and $ \hat{u}_{ij} $ operators  are constant of motions.  Eq.(\ref{u_ham}) 
 reduces to a free Hamiltonian of $\hat{c}_i$  fermions by fixing the bond variables
 $\{u_{ij}=\pm1\}$. The flux configurations ($\{W_p\}$) determine the physics of the spin Hamiltonian (\ref{model_Ham}). However, we have a redundancy to choose the bond variables for a given flux configuration, which reveals the emergent $\mathbb{Z}_2$ gauge field in terms of $u_{ij}$'s.  Since $u_{ij}=-u_{ji}$, we choose the convention to fix $u_{ij}$, where the first index $i$ points to odd sublattices (A,C,E) and the second one refers to even sublattices (B,D,F). Finally, the Hamiltonian
 $ \tilde{H}_{\hat{u},\hat{c}} $ is written in the following compact form    after fixing the gauge configuration,
 \begin{figure*}[!htb]
 	\centering
 	\includegraphics[scale=0.5]{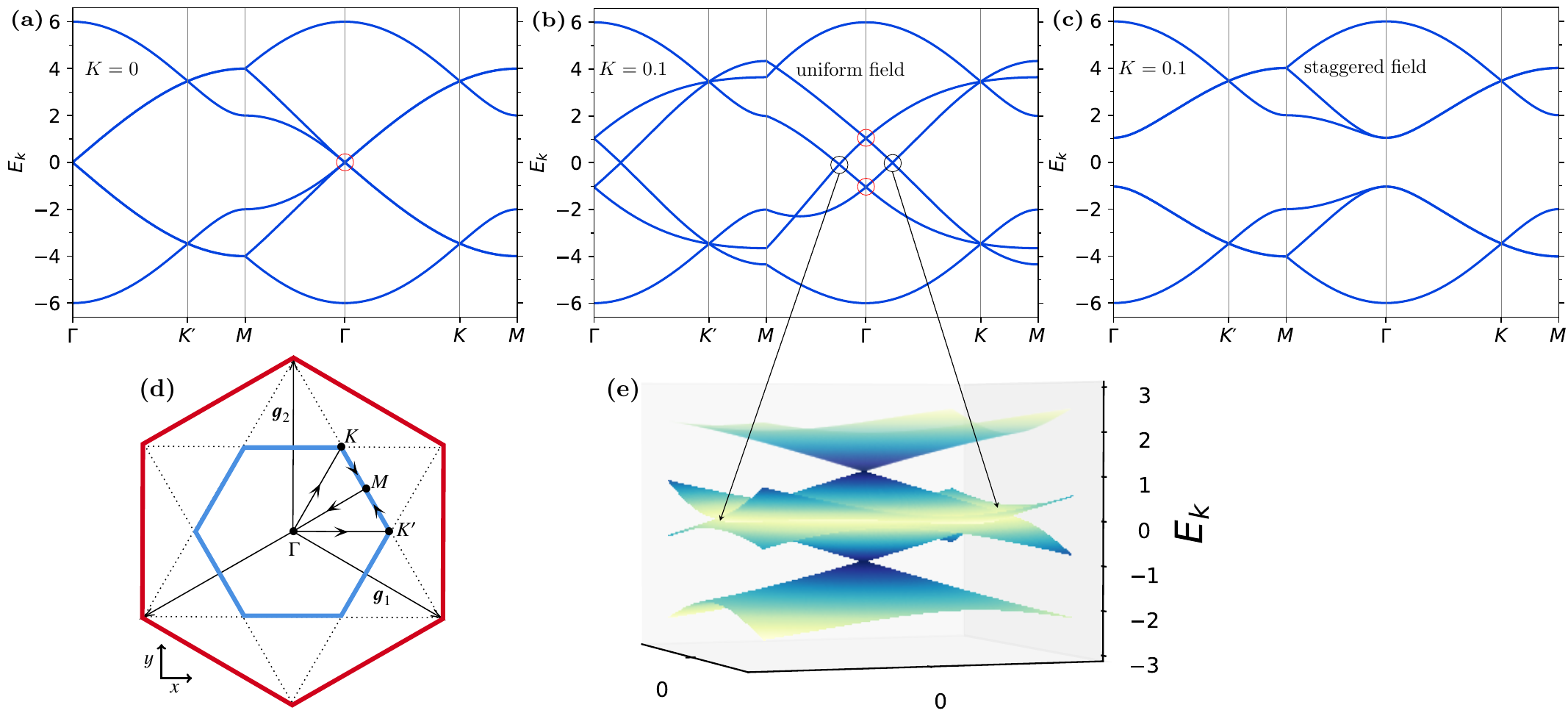}
 	\caption{
 	 The band structure of Kekulé-Kitaev model. (a)  Two degenerate Dirac cones (red circles) are presented at the $\Gamma$ point. (b) The band structure in the presence of a uniform magnetic field, where the Dirac cones remain gapless but non-degenerate. The band crossings marked by the black circles are not Dirac crossing, as shown in (e). 
(c) A staggered magnetic field opens a gap in the Dirac cones.
(d) The paths for presenting the band structure in the reciprocal lattice.
(e) The three-dimensional plot of the Dirac cones in the presence of a uniform magnetic field. 
 	}
 	\label{energy_bands}
 \end{figure*}
\begin{equation}
	\label{Ham_compact}
	\tilde{H}_{\{u\},\hat{c}}=\frac{i}{4}
	\begin{pmatrix}
		\hat{c}_{\text{odd}}^T&\hat{c}_{\text{even}}^T
	\end{pmatrix}
	\begin{pmatrix}
		0&M\\		
		-M^T&0									
	\end{pmatrix}
	\begin{pmatrix}
		\hat{c}_{\text{odd}}\\\hat{c}_{\text{even}}
	\end{pmatrix},
\end{equation}
in which,
\begin{equation}
	\hat{c}_{\text{odd}}=
	\begin{pmatrix}
		\hat{c}_{A}\\	\hat{c}_{C}\\	\hat{c}_{E}
	\end{pmatrix},\quad
	\hat{c}_{\text{even}}=
	\begin{pmatrix}
		\hat{c}_{B}\\	\hat{c}_{D}\\	\hat{c}_{F}
	\end{pmatrix},\quad
M=\begin{pmatrix}
	AB&AD&AF\\
	CB&CD&CF\\
	EB&ED&EF
\end{pmatrix},
\end{equation}
where $ \hat{c}_{\text{odd}/\text{odd}} $ is a $3N$ column vector of  odd/even Majorana operators.
$ M $ is a $ 3N\times 3N $ matrix, whose elements are of the form $ 2J_{\alpha}u_{\langle ij\rangle_{\alpha}} $. The other elements like
$ AB $ is an  $ N\times N $  matrix representing the hopping from A-sublattice to B-sublattice.
The even/odd arrangement considered for the  Majorana operators simplifies the complexity of finding the eigen-energies, because the compact form of Hamiltonian (\ref{Ham_compact}) is 
diagonalized by singular value decomposition (SVD) of $ M_{3N\times 3N} $. 
For other arrangement of Majorana operators one has to diagonalize the whole $ 6N\times 6N $ Hamiltonian.\\
The SVD  of $ M=U S V^T $, where
 $ S=\text{diag}(\varepsilon_1,\dots,\varepsilon_{3N}) $ leads to the diagonalized form
 of Hamiltonian (see Refs. [\onlinecite{Zschocke_Vojta}] and [\onlinecite{choi1}] for more details),
\begin{equation}
	\label{diagonal_ham_pure}
	\tilde{H}_{\{u\},\hat{c}}=\sum_m \varepsilon_m\hat{a}^{\dagger}_m\hat{a}_m-\frac{1}{2}\sum_m \varepsilon_m.
\end{equation}
The complex fermion operators $\hat{a}^{\dagger}_m, \hat{a}_m$ are related to Majorana fermions by the following relations
\begin{equation}
	\label{c_a_relation}
	\begin{pmatrix}
		\hat{c}_{\text{odd}}\\
		\hat{c}_{\text{even}}
	\end{pmatrix}=
	\begin{pmatrix}
		U&U\\
		-iV&iV
	\end{pmatrix}
	\begin{pmatrix}
		\hat{a}\\
		\hat{a}^{\dagger}
	\end{pmatrix}.
\end{equation}

The spin representation in terms of Majorana operators (\ref{partons}) extends the Hilbert space of the model. Hence, an eigenstate of the Hamiltonian (\ref{u_ham}) is composed of two components: gauge and matter, as follows,
	\begin{equation}
		\label{gaugeMatter}
		|\Psi_{\hat{u},\hat{c}}\rangle=|\Psi_{\text{gauge}}^{\hat{u}}\rangle|\Psi_{\text{matter}}^{\{u\},\hat{c}}\rangle,
	\end{equation}
	where,
	\begin{equation}
		\begin{aligned}
			&|\Psi_{\text{gauge}}^{\hat{u}}\rangle=\prod_{\langle jk\rangle_{\alpha}}(\hat{\chi}_{\langle jk\rangle_{\alpha}}^{\dagger})^{n_{\langle jk\rangle_{\alpha}}}|\mathcal{G}\rangle,\quad n_{\langle jk\rangle_{\alpha}}=0,1,\\
			&|\Psi_{\text{matter}}^{\{u\},\hat{c}}\rangle=\prod_{m=1}^{3N}(\hat{a}_{m}^{\dagger})^{n_m}|\mathcal{M}\rangle,\quad n_{m}=0,1,
		\end{aligned}
	\end{equation}
in which $|\mathcal{G}\rangle $ is the vacuum of gauge sectors and $n_{\langle jk\rangle_{\alpha}} $ is the occupation number. 
	$ \hat{\chi}_{\langle jk\rangle_{\alpha}}^{\dagger}$
	is  a complex fermion defined on the lattice bonds through the following relation\cite{Baskaran2007,Knolle_Chalker_Moessner2015,Negahdari_Langari_2023}
	\begin{equation}
		\hat{\chi}_{\langle jk\rangle_{\alpha}}^{\dagger}=\frac{1}{2}(\hat{b}_j^{\alpha}+i\hat{b}_k^{\alpha}),\quad j\in\text{odd},\ k\in\text{even}.
	\end{equation}
	Any arbitrary gauge configuration $\{u\}$ can be determined in terms of the occupation number of $\hat{\chi}$ fermions, using the relation $ \hat{u}_{\langle jk\rangle_{\alpha}}=1-2\hat{\chi}_{\langle jk\rangle_{\alpha}}^{\dagger}\hat{\chi}_{\langle jk\rangle_{\alpha}}^{}$.
	In (\ref{gaugeMatter}), 
	$|\Psi_{\text{matter}}^{\{u\},\hat{c}}\rangle$ is the eigenstate of Hamiltonian (\ref{diagonal_ham_pure}), where $|\mathcal{M}\rangle$ is the vacuum of matter sector in the fixed gauge configuration $\{u\}$. The parity of matter excitations in $ |\Psi_{\text{matter}}^{\{u\},\hat{c}}\rangle$ is fixed, which means it can either be even or odd. For even parity, $ |\Psi_{\text{matter}}^{\{u\},\hat{c}}\rangle$ includes $0, 2, 4, \cdots$-matter fermion excitations,  and for odd parity, it includes $1, 3, 5, \cdots$-matter fermion excitations.

The $0$-flux sector has translational symmetry, which allow us to obtain the dispersion of energy versus lattice momentum, i.e. the band structure.
With the following Fourier transformation,
\begin{equation}
	\hat{c}_{\boldsymbol{k},\alpha}=\frac{1}{\sqrt{N}}\sum_{j}e^{i \boldsymbol{k}\cdot \boldsymbol{R}_j}\hat{c}_{j,\alpha},\quad
	 \alpha=A,B,\dots,F;
\end{equation}
the Hamiltonian (\ref{Ham_compact}) is reduced to,
\begin{equation}
	\label{momentumHam}
	\tilde{H}_{\{u\},\hat{c}}(\boldsymbol{k})=\frac{1}{4}\sum_{\boldsymbol{k}}
	\begin{pmatrix}
		\hat{c}_{\boldsymbol{k},\text{odd}}&
		\hat{c}_{\boldsymbol{k},\text{even}}
	\end{pmatrix}
	\begin{pmatrix}
		0& iM_{\boldsymbol{k}}\\
		- iM_{\boldsymbol{k}}^{\dagger}&0
	\end{pmatrix}
	\begin{pmatrix}
		\hat{c}_{-\boldsymbol{k},\text{odd}}\\
		\hat{c}_{-\boldsymbol{k},\text{even}}
	\end{pmatrix},
\end{equation}
where,
\begin{equation}
\begin{aligned}
	M_{\boldsymbol{k}}=2
	\begin{pmatrix}
		J_z&J_x e^{i\boldsymbol{k}\cdot\boldsymbol{a}}& J_y\\
		J_y&J_z&J_x e^{i\boldsymbol{k}\cdot\boldsymbol{b}}\\
		J_x e^{i\boldsymbol{k}\cdot\boldsymbol{c}}&J_y&J_z
	\end{pmatrix}
\end{aligned}\quad ,\quad 
\begin{aligned}
&\boldsymbol{a}=-\boldsymbol{a}_2\\
&\boldsymbol{b}=\boldsymbol{a}_1\\
&\boldsymbol{c}=\boldsymbol{a}_2-\boldsymbol{a}_1
\end{aligned},
\end{equation}
and $\boldsymbol{a}_1$ and $\boldsymbol{a}_2$ are primitive lattice vectors shown in Fig.\ref{lattice}.
The band structure for the isotropic couplings in the $0$-flux sector is plotted in  Fig.~\ref{energy_bands}(a), which show two degenerate gapless Dirac cones at the $\Gamma$ point. In appendix \ref{energyBands}, we analytically derive the energy-momentum relations.



\subsection{Projection to physical subspace}
\begin{figure*}[!htb]
	\centering
	\includegraphics[scale=1.45]{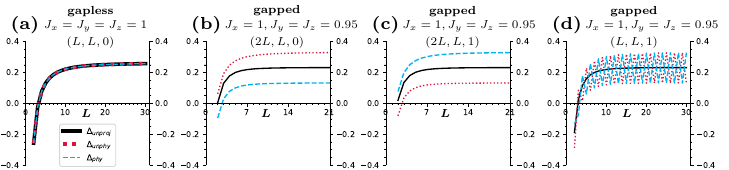}
	\caption{
		The energy difference (denoted by $\Delta$) of the lowest eigenstate in the $0$-flux and $2$-flux sectors, where
		$ \Delta_{\text{unproj}}$ is depicted by solid black line, $\Delta_{\text{unphys}}$ with dotted red line and $\Delta_{\text{phys}}$ by dashed blue line.
		(a) shows the energy gap at the isotropic couplings. (b,c,d) show the same energy gap at 
		$J_z=1$ and $J_y=J_x=0.95$ in gapped phase for different geometries ($L_1, L_2, M$) presented in each panel.  Note that in (c) and (d), the parameter $M$ is non-zero. As shown in (b), (c) and (d), in the gapped phase, the difference between physical and unphysical data persists up to the thermodynamic limit, similar to the Kitaev model\cite{Pedrocchi_Loss}.
	}
	\label{phy_unphy_unproj}
\end{figure*}

Although representation (\ref{partons}) makes the problem solvable, it also expands the Hilbert space to $4^{6N}$ instead of $2^{6N}$, which includes the unphysical sectors of gauge fields. Using the following projection operator, we can determine the physical Hilbert space. Suppose $ |\Psi_{\hat{u},\hat{c}}\rangle $ is an eigenvector of  $ \tilde{H}_{\hat{u},\hat{c}} $, the physical state is obtained by the following projection,
\begin{align}
	|\Psi_{\text{phys}}\rangle=\hat{\mathcal{P}}|\Psi_{\hat{u},\hat{c}}\rangle ,\qquad
	\hat{\mathcal{P}}= \hat{\mathcal{S}} \hat{\mathcal{P}}_0,
\end{align}
where 
$  \hat{\mathcal{P}}_0=\frac{1+\hat{D}}{2} $, and 
$\hat{D}=\prod_{i=1}^{6N}\hat{b}_i^{x}\hat{b}_i^{y}\hat{b}_i^{z}\hat{c}_i$\cite{Pedrocchi_Loss,Yao_Kivelson_2009}. If $ D=+1 $, then  $ |\Psi_{\hat{u},\hat{c}}\rangle $ is a physical state. $ \hat{D} $ can be written in the following simple and compact form in term of the parities of the gauge ($\hat{\pi}_{\chi}$) and matter ($\hat{\pi}_a$) fermions (see appendix \ref{D_operator}), 
\begin{equation}
	\label{parity_proj}
	\hat{D}=(-1)^{\theta}\det(\Lambda_u)\hat{\pi}_{\chi}\hat{\pi}_a,\quad
	\text{with} \quad\theta=L_1-M-1,
\end{equation}
where $ \Lambda_u $ is obtained from the diagonalization of the Hamiltonian (\ref{Ham_compact})\cite{Pedrocchi_Loss,Zschocke_Vojta,Negahdari_Langari_2023}. Note that the  factor $\theta$ differs from the corresponding one in the Kitaev model\cite{Pedrocchi_Loss}. Since $ \det(\Lambda_u)\hat{\pi}_{\chi} $ is  gauge invariant, the parity of the physical state $ |\Psi_u\rangle  $ in each flux sector is obtained by the following equation,
\begin{equation}
	\pi_a=\frac{(-1)^{\theta}}{	\det(\Lambda_u)\pi_{\chi}}.
\end{equation}
Fig.~\ref{phy_unphy_unproj} shows the energy difference of the lowest states in the 0-flux and 2-flux sectors for different boundary conditions in the unprojected space (i.e., the extended Hilbert space) and projected subspaces (i.e., physical and unphysical subspaces). In the unprojected space, this energy difference is equal to the two-vison gap ($ \Delta_v $),
\begin{eqnarray}
	\label{eq_20_chp3}
	\Delta_{\text{unproj}}=E_2-E_0=\Delta_v
\end{eqnarray}
where $  E_0$ and $ E_2 $ are the last term ($-\sum_m \frac{\varepsilon_m}{2}$) in Eq.(\ref{diagonal_ham_pure}) for the $0$- and $2$-flux sectors, respectively. The energy differences in the projected subspaces is defined such that: if the parity of matter fermions in the physical subspace of the 0- and 2-flux is equal to $+1$ and $-1$, respectively, we consider the reverse parities $-1$ and $+1$ for them in the unphysical subspace. So,
\begin{equation}
	\label{eq_21_chp3}
	\begin{aligned}
		&\Delta_{\text{phys}}=(E_2+\bar{\varepsilon}_1)-E_0=\Delta_v+\bar{\varepsilon}_1\\
		&\Delta_{\text{unphys}}=E_2-(E_0+\varepsilon_1)=\Delta_v-\varepsilon_1
	\end{aligned}
\end{equation}
where $ \varepsilon_1 $ and $ \bar{\varepsilon}_1 $ are the lowest matter excitation in  the $ 0$- and $2$-flux sectors, respectively. The aforementioned  flux sectors are gapless at the isotropic coupling $J_x=J_y=J_z=J $, which leads to equal energy difference within all subspaces, 
$\Delta_v \approx 0.26J$, as shown in Fig.~\ref{phy_unphy_unproj}(a). 
For small $L$, all parts in Fig.~\ref{phy_unphy_unproj} indicate that the energy of the $2$-flux state is lower than the $0$-flux state. This observation appears to be inconsistent with Lieb’s theorem\cite{Lieb_flux_1994}. Similar behavior is also seen in the Kitaev model\cite{Zschocke_Vojta}. 
Notably, while the original Hamiltonian has been transformed into a free fermion Hamiltonian using Kitaev parton construction, the resulting Hamiltonian is more than just a free fermion problem due to the extension of Hilbert space, where we have to consider the effect of the projection operator. Indeed, the conditions of Lieb's theorem are not fully satisfied, since the projection operator may introduce excitations to the free fermionic states, leading to an increase in energy.
Nevertheless, the ground state belongs to the $0$-flux sector for large system sizes.   

The Kekulé-Kitaev  model has two topological Wilson loops on a torus, characterized by eigenvalues $l_1=\pm1$ and $l_2=\pm1$\cite{Halasz_Chalker_2014}. In each flux sector $\{W_p\}$, there exist four distinct topological states denoted as  $|\{W_p\},l_1,l_2\rangle$. The dotted links in Fig.~\ref{topological_loop} illustrate  the  Ising terms that constitute these topological operators along  directions $\boldsymbol{a}_1 $ and $\boldsymbol{a}_2 $ for the  geometry $L_1=3$, $ L_2=2$, and $M=0 $. Similar examples can be found in  Ref.[\onlinecite{Negahdari_Langari_2023}]. In the gapless phase of the Kitaev model, the physical state within the $0$-flux sector, which is characterized by the topological labels $ (l_1=+1,l_2=+1) $ has an odd parity $ \pi_a $  for all sets  of  $(L_1, L_2,M)$\cite{Zschocke_Vojta}. The odd parity constraint also exists in the non-Abelian phase of the Kitaev model\cite{Negahdari_Langari_2023}. However,  we numerically observed that the odd parity constraint does not hold in the gapless phase of the Kekulé-Kitaev model.

\subsection{Effects of weak uniform and staggered  magnetic fields}\label{uniform_and_staggered}

\begin{figure}[!h]
	\includegraphics[width=1\columnwidth]{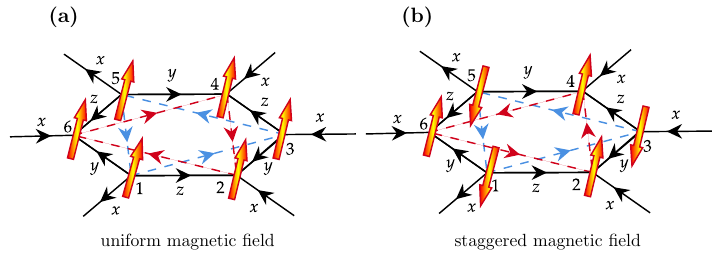}
	\caption{
		Black arrows represent the nearest neighbor hopping with  $u_{ij}=+1$. (a) The second neighbor hoppings  (indicated by red and blue dashed lines) exhibit opposite rotational direction  in the presence of a uniform magnetic field.  (b) However, in the presence of a staggered magnetic field, the second neighbors exhibit the same rotations. The staggered magnetic field may be realized by proximity to an antiferromagnetic honeycomb lattice material. The up and down arrows indicate  the local magnetic fields $(h_x,h_y,h_z)$ and  $(-h_x,-h_y,-h_z)$, respectively.
	}
	\label{kekule_kitave_second_hop}
\end{figure}

We consider the effect of both uniform and staggered weak magnetic fields, as depicted  in Fig.~\ref{kekule_kitave_second_hop}, 
on the Kekulé-Kitaev model at the isotropic couplings. 
The effect of magnetic field has the following simple form
\begin{align}
	&V=-\sum_{\text{i}}(h_x\hat{\sigma}_i^x+h_y\hat{\sigma}_i^y+h_z\hat{\sigma}_i^z).
\end{align}
 The dominant contribution of the weak magnetic field within a perturbative approach is given by the following three-spin term,
\begin{equation}
	\label{model_Ham_magnetic}
	\hat{H}_{\text{eff}}=\hat{H}-K\sum_{\substack{\langle ik\rangle_{\alpha},\langle kj\rangle_{\beta}\\\gamma\perp\alpha,\beta}}\hat{\sigma}^{\alpha}_i\hat{\sigma}^{\gamma}_k\hat{\sigma}^{\beta}_j \;\;\;\; ; \;\;\;\; K\propto6\frac{h_xh_yh_z}{\Delta_v^2},
\end{equation}
where the factor  $6 = 3!$ arises from different permutation of spins, and $\Delta_v$ is the two-flux gap (or two-vison gap). According to the plaquette depicted in Fig.~\ref{kekule_kitave_second_hop}, the following three-spin interactions represent the effect
of magnetic field
\begin{equation}
	\label{three_spin}
	\begin{aligned}
		&+\hat{\sigma}_1^{z}\hat{\sigma}_3^{y}\hat{\sigma}_2^{x}+
		\hat{\sigma}_3^{z}\hat{\sigma}_5^{y}\hat{\sigma}_4^{x}+
		\hat{\sigma}_5^{z}\hat{\sigma}_1^{y}\hat{\sigma}_6^{x}\\
		&\pm\hat{\sigma}_2^{z}\hat{\sigma}_6^{y}\hat{\sigma}_1^{x}
		\pm
		\hat{\sigma}_6^{z}\hat{\sigma}_4^{y}\hat{\sigma}_5^{x}\pm
		\hat{\sigma}_4^{z}\hat{\sigma}_2^{y}\hat{\sigma}_3^{x},
	\end{aligned}
\end{equation}
where the plus and minus signs in the second line of Eq.(\ref{three_spin}) correspond to the uniform and staggered magnetic fields, respectively.

The three-spin interactions are transformed to the next-nearest neighbor hopping of Majorana fermions in the physical subspace as given by\cite{Hermanns2015}
\begin{equation}
\hat{\sigma}_j^{\alpha}\hat{\sigma}_k^{\beta}\hat{\sigma}_l^{\gamma}=-i\epsilon^{\alpha\beta\gamma}u_{\langle jl\rangle_{\alpha}}u_{\langle kl\rangle_{\beta}}\hat{c}_j \hat{c}_k.	
\end{equation}
Therefore, we  generalize Eq.(\ref{Ham_compact}) to the following equation,
\begin{equation}
	\label{Ham_compact_mag}
	\tilde{H}_{\text{eff}}^{\{u\},\hat{c}} =\frac{i}{2}
	\begin{pmatrix}
		\hat{c}_{\text{odd}}^T&\hat{c}_{\text{even}}^T
	\end{pmatrix}
	\begin{pmatrix}
		F_o&M\\		
		-M^T&-F_e								
	\end{pmatrix}
	\begin{pmatrix}
		\hat{c}_{\text{odd}}\\\hat{c}_{\text{even}}
	\end{pmatrix},
\end{equation}
where $ F_o $ and $ F_e $ represent the next-nearest neighbor hopping matrices in odd and even sublattices. Using the following complex fermions,

\begin{figure}[!t]
	\includegraphics[width=0.5\columnwidth]{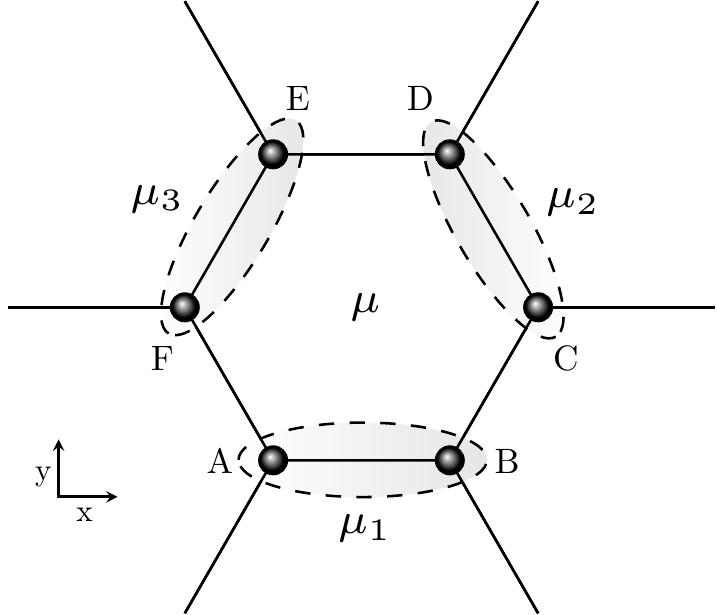}
	\caption{In the unit cell $\mu$, there are three z-bonds labeled $\mu_1$, $\mu_2$, and $\mu_3$, each hosting complex fermions $\hat{f}_1$, $\hat{f}_2$, and $\hat{f}_3$ on them, respectively.	}
	\label{Majorana_paring}
\end{figure}

\begin{equation}
	\begin{aligned}
		\hat{f}_1=&\frac{1}{2}(\hat{c}_A+i\hat{c}_B),\\
		\hat{f}_2=&\frac{1}{2}(\hat{c}_C+i\hat{c}_D),\\
		\hat{f}_3=&\frac{1}{2}(\hat{c}_E+i\hat{c}_F),
	\end{aligned}\quad \text{or} \quad
	\hat{f}=\frac{1}{2}(\hat{c}_{\text{odd}}+i\hat{c}_{\text{even}}),
\end{equation}
which are defined on AB, CD, and EF links, respectively (see Fig.~\ref{Majorana_paring}), the Hamiltonian $ 	\tilde{H}_{\text{eff}}^{\{u\},\hat{c}} $ is  written in the following form
\begin{equation}
	\label{Ham_compact_mag_complex}
	\tilde{H}_{\text{eff}}^{\{u\},\hat{c}} =\frac{1}{2}\begin{pmatrix} 
		\hat{f}^{\dagger}&\hat{f}\end{pmatrix}
	\begin{pmatrix} 
		\tilde{h}&\tilde{\Delta}\\
		\tilde{\Delta}^{\dagger}& -\tilde{h}^{*}
	\end{pmatrix}
	\begin{pmatrix} 
		\hat{f}\\\hat{f}^{\dagger}\end{pmatrix},
\end{equation}
where,
\begin{align}
	\tilde{h}&=(M^T+M)+i(F_o-F_e),\quad
	\tilde{h}^{\dagger}=\tilde{h},
	\nonumber\\
	\tilde{\Delta}&=(M^T-M)+i(F_o+F_e),\quad \tilde{\Delta}^T=-\tilde{\Delta}.
\end{align}
The Hamiltonian (\ref{Ham_compact_mag_complex}) can be diagonalized using a unitary transformation, as given in Ref. [\onlinecite{Knolle_Chalker_Moessner2015}]. 
Fig.~\ref{energy_bands}(b) shows the energy bands of the Kekulé-Kitaev  model in the presence of a uniform magnetic field within the $0$-flux sector, where the Dirac cones remain massless similar to the zero field case, Fig.~\ref{energy_bands}(a). Note that the band crossing indicated by the black circles in Fig.~\ref{energy_bands}(b) at finite momentum, as shown by the arrows in Fig.~\ref{energy_bands}(e), are not Dirac crossings. 
However, the uniform magnetic field lifts the degeneracy of the two Dirac cones. In contrast, introducing a staggered magnetic field results in the band structure shown in Fig.~\ref{energy_bands}(c), which leads to massive Dirac cones.

\begin{figure}[!t]
	\includegraphics[width=1\columnwidth]{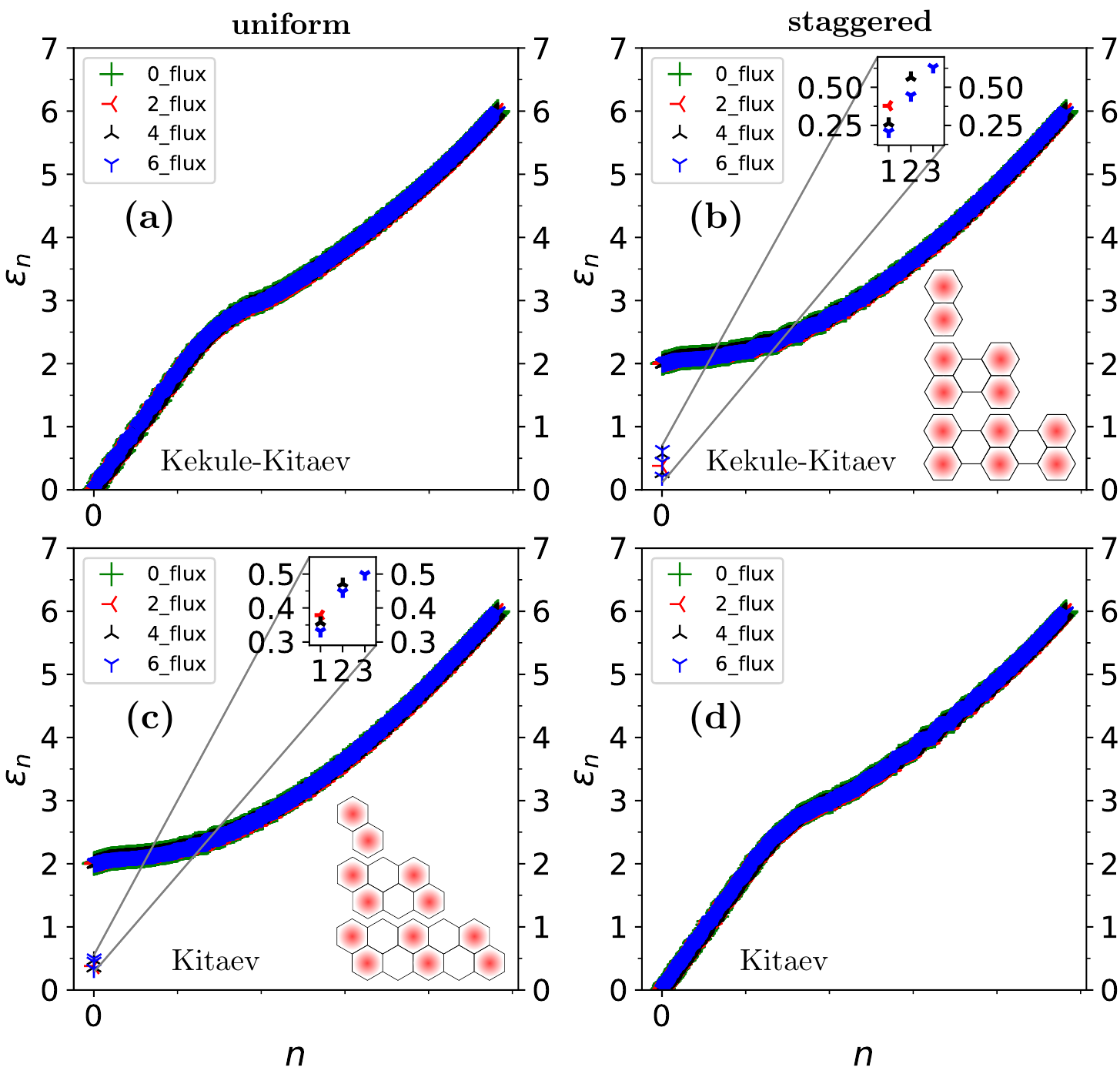}
	\caption{The energy levels ($\varepsilon_{n}$) of the Kekulé-Kitaev model for a system with $L_1=L_2=30$ and $M=0$  exhibit distinct behavior in the presence of (a) uniform and (b) staggered magnetic fields at the isotropic coupling $J=1$ and $K=0.2$. In (a), no Majorana bound state is observed, whereas in  (b), for each $2n$-flux, $n$ Majorana bound states are imposed inside the fermionic gap. The inset in (b) illustrates the $(2, 4, 6)$-flux configurations in our numerical simulation. (c) and (d) illustrate the energy levels of the Kitaev model in both uniform and staggered magnetic fields, respectively. The flux configurations we considered for the Kitaev model are depicted in the inset of (c). Notably, the behavior of Kitaev model in these two types of fields is contrary to that of the Kekulé-Kitaev model.}
	\label{bound_states}
\end{figure}

We have plotted the energy levels ($\varepsilon_{n}$) versus $n$ of both Kekulé-Kitaev and Kitaev models within several flux sectors in Fig.~\ref{bound_states} for both uniform and staggered magnetic fields.
In the presence of a uniform magnetic field (Fig.~\ref{bound_states}(a)) the spectrum of Kekulé-Kitaev model is gapless and there is no bound state contrary to the Kitaev model, Fig.~\ref{bound_states}(c), where the spectrum is gapped and a non-Abelian phase emerges \cite{lahtinen2008,Lahtinen_Interacting_2011,Knolle2016_theses}.
The non-Abelian phase of the Kitaev model has threefold topological degeneracy\cite{Kells_2009,Negahdari_Langari_2023}, which is consistent with the Ising type of the non-Abelian anyons.
For the Kekulé-Kitaev model in a staggered magnetic field (Fig.~\ref{bound_states}(b)), the Majorana bound states are imposed inside the fermionic gap by inserting flux excitations and the non-Abelian statistics of the vortices emerge from these bound states,  while the Kitaev model remains  in a gapless  phase, Fig.~\ref{bound_states}(d). 
We have also numerically observed the threefold topological degeneracy for the Kekulé-Kitaev model in its non-Abelian phase.

\subsection{Stability of gapless Dirac Cones}
\label{Stability}
The Dirac cones at the isotropic coupling of the Kekulé-Kitaev model remain gapless if we apply a uniform magnetic field, while applying a staggered magnetic field makes them massive. Hence, the relevant question is what symmetry or symmetries are responsible for the stability of the gapless Dirac cones at the isotropic coupling?
	
Gapless Dirac cones exist only at the isotropic coupling of the Kekulé-Kitaev model. At this coupling, in addition to the general symmetries of model\cite{Kekule_Kitaev_mossner}, there exists an extra symmetry. The latter symmetry consists of a simultaneous combination of 
 inversion (or a $\pi$ rotation of the lattice around the center of any hexagonal plaquette $W_{\alpha}$) and a global spin rotation by the angle 
 $\pi/2$ around the $\alpha$-axis, where $\alpha=x,y,z$. Let us denote this symmetry by $\hat{I}$. In general\cite{Bernevig2013}, a necessary condition for Dirac cones to be gapless is the Hamiltonian being invariant under the product of time reversal ($\hat{\tau}$) and space inversion ($\hat{I}$) symmetry, i.e., $\hat{\tau}\hat{I}$.  It can be straightforwardly examined that the original Kekulé-Kitaev model and the corresponding effective model in the presence of a uniform magnetic field possess $\hat{\tau}\hat{I}$ symmetry. However, this symmetry is broken in the effective model in the presence of a staggered magnetic field. Consequently, as shown in Fig.~\ref{energy_bands}(c), Dirac cones become gapped by adding a staggered field. As previously stated, note that $\hat{I}$ includes global spin rotations as well. The presence or absence of $\hat{\tau}\hat{I} $ symmetry can be analyzed in the spin representation of both Hamiltonians (i.e., Eq.(\ref{model_Ham}) and Eq.\ref{model_Ham_magnetic})) and also in terms of Majorana fermion representations. Appendix \ref{StabilityOfDirac} provides a detailed explanation of how this symmetry is examined in the Majorana fermion representation.

\section{Linear response: dynamical spin structure factor}\label{linear_response_sec}
 \begin{figure}[!t]
	\includegraphics[width=0.93\columnwidth]{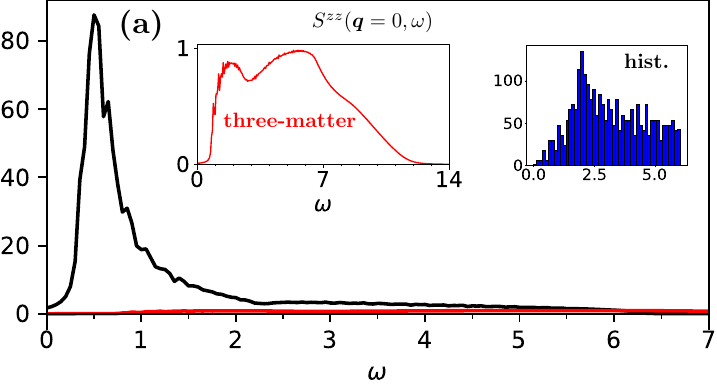}
	\includegraphics[width=0.93\columnwidth]{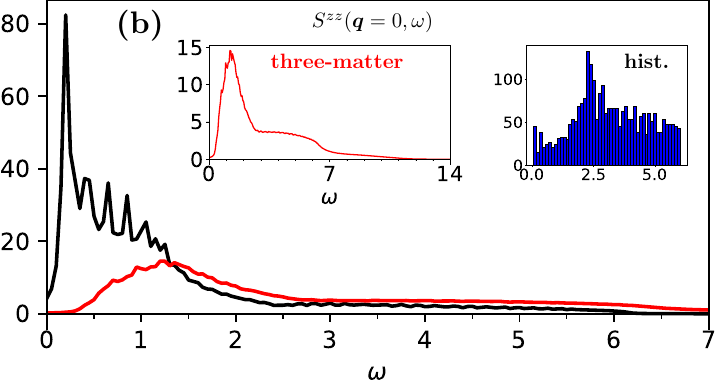}
	\includegraphics[width=0.93\columnwidth]{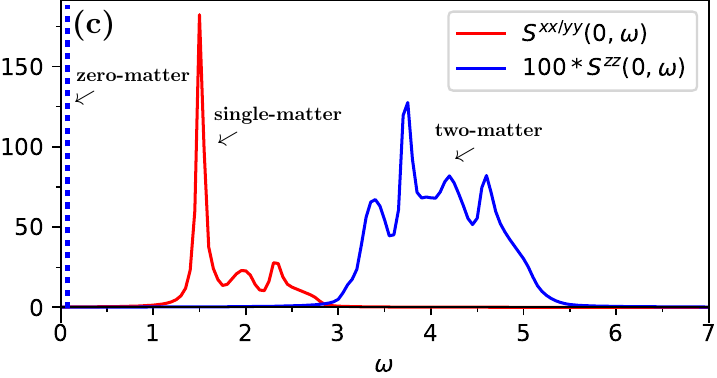}
	\caption{
		(a) The contributions of single-matter fermion (black line) and three-matter fermion excitations (red line) of $S^{zz}(0,\omega)$  in the Kekulé-Kitaev model and (b) in the presence of a uniform magnetic field with $K=0.1$ and isotropic couplings. Moreover, the histograms of energy levels, which correspond to the density of states, are provided in the right inset.  In the middle insets, the contribution of three-matter fermion excitations is shown more clearly. (c) shows $S^{xx}(0,\omega)$ and $S^{yy}(0,\omega)$ within the single-matter fermion approximation, as well as the $S^{zz}(0,\omega)$  in the zero- and two-matter fermion approximations without any magnetic field for the couplings $J_z=1, J_y=J_x=0.2$.  The dashed line, which represents a delta function at $\omega=0.011$, indicates the contribution of the zero-matter fermion excitation. Tiny numerous peaks observed in panels (a) and (b) do not have any specific physical significance, which arise from finite-size effects. In these calculations, a system with $30\times30$ unit cells was considered.
	}
	\label{stacture_factor_q0}
\end{figure} 
The dynamical spin structure factor (DSSF) is an experimentally significant quantity, as it is related to the cross-section in elastic neutron scattering experiments. The total DSSF is given with the following equation\cite{Knolle_Chalker_Moessner2014,Knolle_Chalker_Moessner2015},
\begin{equation}
	\label{stracture_factor}
	S(\boldsymbol{q},\omega)=\frac{1}{N}\sum_{\boldsymbol{R}_j,\boldsymbol{R}_k}\sum_{\alpha,\beta}
	e^{-i\boldsymbol{q}\cdot(\boldsymbol{R}_j-\boldsymbol{R}_k)}\int_{-\infty}^{+\infty}dt e^{i\omega t}
	S^{\alpha\beta}_{jk}(t).
\end{equation}
Our focus is on the zero temperature response, where we denote the ground state by $|G\rangle$ and,
\begin{equation}
	\label{correlation}
	S^{\alpha\beta}_{jk}(t)=\langle G| \hat{\sigma}_j^{\alpha}(t)\hat{\sigma}_k^{\beta}(0)|G\rangle.
\end{equation}
The correlation function in Eq.(\ref{correlation}) is extremely short-ranged, where only on-site and the nearest neighbors correlations are non-zero\cite{Baskaran2007}. We write the DSSF in the following Lehmann representation,
\begin{equation}
	\label{lehmann}
	S(\boldsymbol{q},\omega)=2\pi \sum_{\substack{j,k,\\\alpha,P}} e^{-i\boldsymbol{q}\cdot\boldsymbol{R}_{kj}}
	\delta(\omega+E_0-E_P)
	\langle G|\hat{\sigma}_j^{\alpha}|P\rangle 
	\langle P|\hat{\sigma}_k^{\alpha}|G\rangle,
\end{equation}
where $j$ and $k$ are restricted to a single unit cell. Given that the ground state belongs to the zero-flux sector, $ |P\rangle $  state in Eq.(\ref{lehmann}) belongs to the two-adjacent flux sector. If we denote the energy of the matter excitations in $ |P\rangle $ state by $ \varepsilon_{p} $, the DSSF has peaks at the frequencies $ \omega=\Delta_v+\varepsilon_{p} $.

All matter fermion excitations from few-matter fermion to many-matter fermion excitations
are of the order $\mathcal{O}(2^{3N})$ contribute to the summation in $(\ref{lehmann}) $. Since we are examining the system’s response at zero temperature, it is reasonable to expect that few-matter fermion excitations significantly influence the overall response. Both analytical and numerical investigations have confirmed the validity of the single-matter fermion excitation approximation within the context of the Kitaev model. Notably, this approximation effectively captures the system’s response at zero temperature \cite{Knolle_Chalker_Moessner2014,Knolle_Chalker_Moessner2015,Knolle2016_theses}.

In order to examine the validity of the single-matter fermion approximation in the Kekulé-Kitaev model, we have calculated the  contributions arising from both single-matter fermion and three-matter fermion excitations in $ S^{zz}(\boldsymbol{q}=0,\omega) $ at the isotropic coupling $J_x=J_y=J_z=1$ (see appendix \ref{Matrix_elements} for details). Figs.~\ref{stacture_factor_q0}(a,b) show the results for two cases: the Kekulé-Kitaev model and the model in the presence of a weak uniform magnetic field. The results reveals that the influence of three-matter fermion excitations (and higher orders) in both cases is negligible compared with the contribution of single-matter fermion excitations.
Consequently, the single-matter fermion approximation is well-suited to calculate the structure factor of the Kekulé-Kitaev model. In the presence of a weak uniform magnetic field, the contribution of three-matter fermion  excitations is more significant than their influence without the field. The insets in Figs.~\ref{stacture_factor_q0}(a,b) show that this behavior is attributed to the higher density of states at low energies of the perturbed model.

Digressed from the isotropic coupling to $J_z=1,J_y=J_x=0.2$,   Fig.~\ref{stacture_factor_q0}(c) shows $ S^{xx/yy}(\boldsymbol{q}=0,\omega) $  in the single-matter fermion excitation approximation and $ S^{zz}(\boldsymbol{q}=0,\omega) $ both in the zero- and two-matter fermion  excitations approximation. The zero-matter fermion contribution is represented by a dashed line. This signal corresponds to a delta function with a relatively large amplitude, positioned at the energy difference between 0-flux and 2-flux sectros, specifically at $\Delta=E_2-E_0\approx0.011$, which is a very small value at this coupling. Again, we observe that the two-matter fermion excitations have a tiny contribution to  $ S^{zz}(\boldsymbol{q}=0,\omega) $ compared with the zero-matter fermion excitation, confirming the validity of the few-matter  fermion approximation. Details of the calculation of the contributions from zero- and two-matter fermion excitations are provided in the appendix \ref{Matrix_elements}.

\begin{figure}[!t]
	\includegraphics[width=0.99\columnwidth]{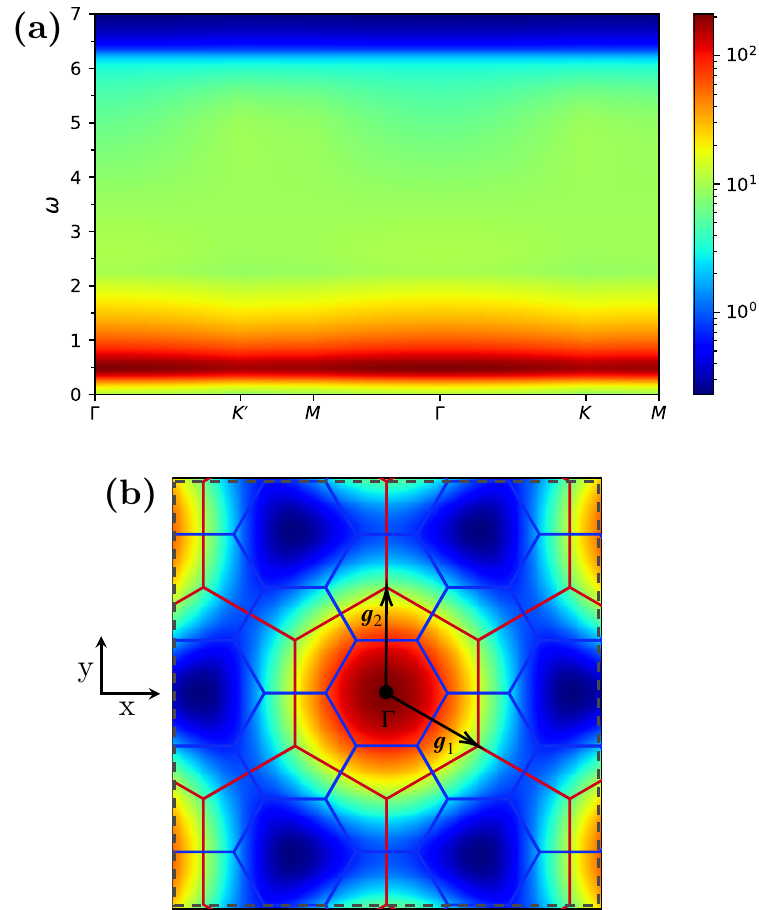}
	\caption{
		(a) The density plot of total DSSF ($S(\boldsymbol{q},\omega)$) along the path $\Gamma K' M\Gamma K M $ at isotropic coupling $J=1$ for the  Kekulé-Kitaev model. At this coupling 
		$ S^{xx}(\boldsymbol{q},\omega)=S^{yy}(\boldsymbol{q},\omega)=S^{zz}(\boldsymbol{q},\omega)$. 
		 (b) The density plot of $S(\boldsymbol{q}, \omega=1)$ in the reciprocal lattice. 
The presence of a broad continuum in both energy and momentum spaces arises from spin fractionalization and the lack of long-range order. These results are for a geometry of $L_1=L_2=30$ and $ M=0.$
	}
	\label{stracture_path_all_BZ}
\end{figure}

Based on the single-matter fermion approximation we have plotted the total DSSF  
$S(\boldsymbol{q},\omega)$ of the Kekulé-Kitaev model at the isotropic coupling in Fig.~\ref{stracture_path_all_BZ}. Panel (a) shows the density plot of $S(\boldsymbol{q},\omega)$ 
along the  path $\Gamma K' M\Gamma K M $, and in panel (b) the density plot is shown in the entire Brillouin zone at fixed $\omega=1$.
Both patterns of the DSSF are qualitatively similar to those reported for the Kitaev model \cite{Knolle_Chalker_Moessner2014,Knolle2016_theses,Banerjee_Jiaqiang_Science_2017}. 
Consequently, both Kekulé-Kitaev and Kitaev models exhibit nearly identical linear responses. 

In Fig.~\ref{stacture_factor_q0} and Fig.~\ref{stracture_path_all_BZ} we have considered a system with $L_1=L_2=30$ and $ M=0$. In order to reduce the finite-size effects, we substituted the delta function in Eq.(\ref{lehmann}) with the Lorentzian function 
\begin{equation}
\delta(x)= \lim_{\gamma \rightarrow 0} \frac{1}{\pi}\frac{\gamma}{x^2+\gamma^2}
\end{equation}
characterized by the broadening parameter  $\gamma=0.04J_z$.

We have also derived the DSSF using the Pfaffian approach, as proposed in Ref.~[\onlinecite{Knolle_Chalker_Moessner2015}], which corroborates our results, although it was not included in this work. Due to computational resource limitations, we analyzed a system of $8\times8$ unit cells using the Pfaffian approach, which accounts for all matter fermion  excitations. The results from this method also confirm that the single-matter fermion approximation is valid for zero-temperature responses. However, the Pfaffian approach considers all matter fermion  excitations, while the parity of these excitations should be odd or even, consistent with the projection to the physical subspace, Eq.(\ref{parity_proj}). Therefore, one might consider how to generalize the Pfaffian approach under odd or even parity constraints.
 
\section{Nonlinear response: two-dimensional coherent spectroscopy}\label{nonlinear_response_sec}
In two-dimensional coherent spectroscopy, the system is exposed to two magnetic  impulses $B(t)$, with a time interval of $ \tau_1 $\cite{Armitage_nonlin,choi1},
\begin{equation}
	\boldsymbol{B}(t)=B_{0}\delta(t)\hat{e}_{z}+B_{1}\delta(t-\tau_1)\hat{e}_{z},
\end{equation}
and then after time $\tau_2$, the magnetization   
$ M^{z}_{01}(\tau_1+\tau_2) $ is recorded. For simplicity, we considered the same linear polarization for these two impulses and the recorded signal. The definition of nonlinear magnetization is:
\begin{equation}
	M^{z}_{NL}(\tau_1+\tau_2)=M_{01}^{z}(\tau_1+\tau_2)-M_{0}^{z}(\tau_1+\tau_2)-M_{1}^{z}(\tau_1+\tau_2), 
	\label{nonlniear_mag}
\end{equation}
where $ M_{0}^{z}(\tau_1+\tau_2) $ and $ M_{1}^{z}(\tau_1+\tau_2) $ are the recorded magnetization as a response to only the single impulse $ B_{0} $ and  $ B_{1} $ at time $ t=\tau_1+\tau_2$, respectively. For weak impulses, we can express the nonlinear magnetization in terms of different orders of nonlinear susceptibilities,\cite{choi1}
\begin{align}
	\label{non_lin_formula}
	M^{z}_{NL}(\tau_1+\tau_2)/N_{\text{site}}&=B_{0}B_{1}\chi^{(2),z}_{zz}(\tau_2,\tau_1)\nonumber\\
	&+B_{1}B_{0}B_{0}\chi^{(3),z}_{zzz}(\tau_2,\tau_1,0)\nonumber\\
	&+B_{1}B_{1}B_{0}\chi^{(3),z}_{zzz}(\tau_2,0,\tau_1)+\mathcal{O}(B^4).
\end{align}
The second-order susceptibility is zero\cite{choi1}, and as a result, the first non-zero contributions to non-linear response are the third-order susceptibilities, which are given with the following relations,
\begin{figure*}[!htb]
	\centering
	\includegraphics[scale=0.47]{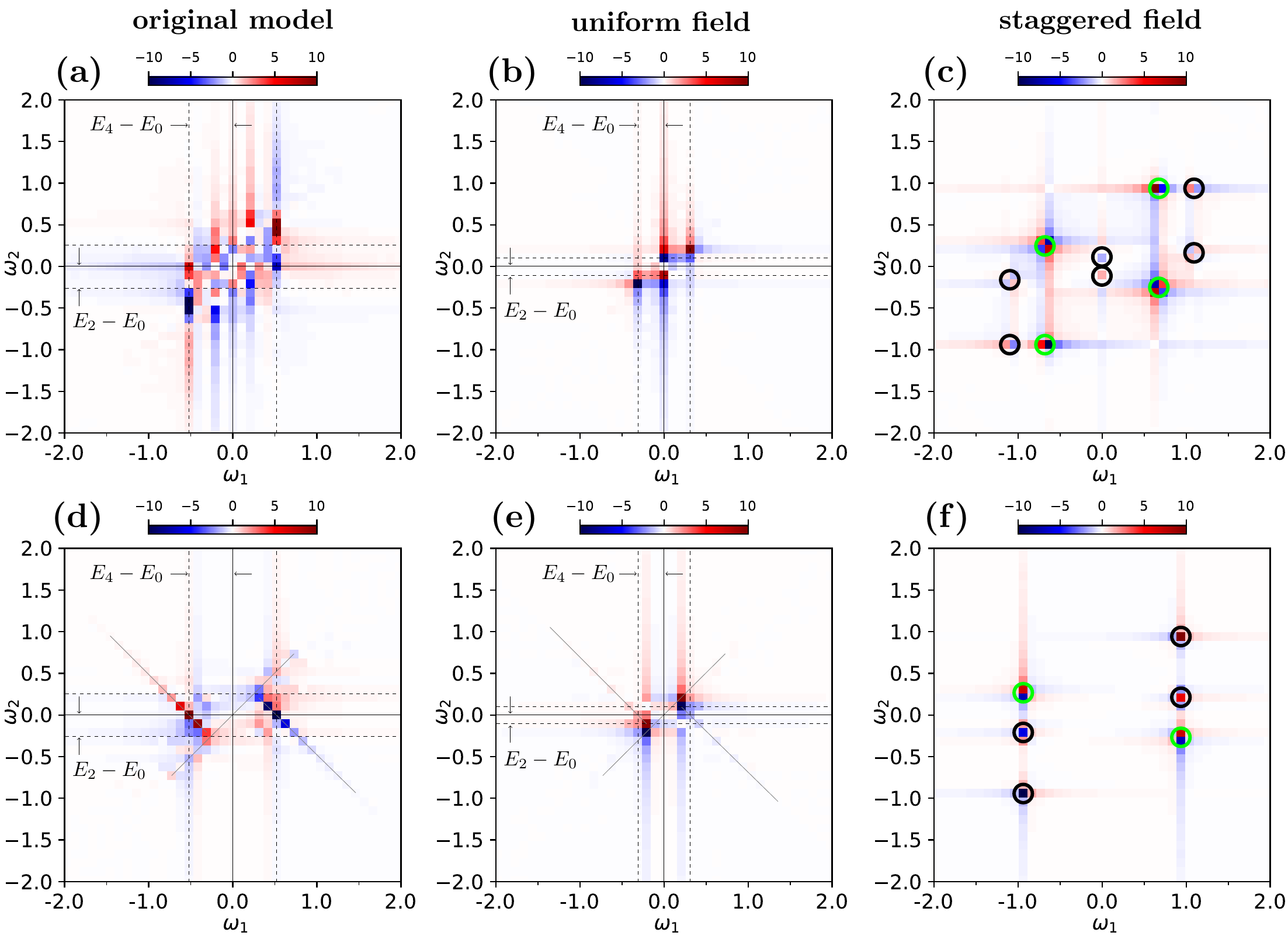}
	\caption{(a,d) Fourier transforms of the two-dimensional non-linear susceptibilities of the original Kekulé-Kitaev model, (b,e) the model in the presence of a weak uniform magnetic field with $K=0.1$ and (c,f) the model in the presence of a weak staggered  magnetic field with $K=0.2$.  All models are evaluated at isotropic couplings  $(J_x=J_y=J_z=1) $. Panels  (a, b, c) show $\chi^{(3),z}_{zzz}(\omega_2,\omega_1,0)$, while panels  (d, e, f) show $\chi^{(3),z}_{zzz}(\omega_2,0,\omega_1)$. In panels (c) and (f), the signals marked with green circles originate from two non-adjacent fluxes, while those marked with black circles arise from other flux configurations. All responses are normalized to 10.}
	\label{nonlinear_response}
\end{figure*}
\begin{align}
	\chi^{(3),z}_{zzz}(\tau_2,\tau_1,0)&=\frac{2}{N_{\text{site}}}\sum_{l=1}^{4}\Im\bigg[\Upsilon^{(l),z}_{zzz}(\tau_2,\tau_1,0)\bigg],\\
	\chi^{(3),z}_{zzz}(\tau_2,0,\tau_1)&=\frac{2}{N_{\text{site}}}\sum_{l=1}^{4}\Im\bigg[\Upsilon^{(l),z}_{zzz}(\tau_2,0,\tau_1)\bigg].
\end{align}
$ \Upsilon^{(l),z}_{zzz} $ is the four-point correlation function at zero temperature.
For instance,
\begin{align}
	\label{R_12_general}
	&\Upsilon^{(1),z}_{zzz}(\tau_2,\tau_1,0)=\langle \hat{M}_I^{z}(0)\hat{M}_I^{z}(\tau_1)\hat{M}_I^{z}(\tau_1+\tau_2)\hat{M}_I^{z}(0)\rangle\nonumber\\
	&=\sum_{\mu_i\nu_j\lambda_k\rho_l}\sum_{PQR}\langle G|\hat{Z}_{\mu_i}|P\rangle \langle P|\hat{Z}_{\nu_j}|Q\rangle \langle Q|\hat{Z}_{\lambda_k}|R\rangle \langle R|\hat{Z}_{\rho_l}|G\rangle\nonumber\\
	&\quad\quad\quad\quad\times e^{i(E_P-E_R)\tau_1+i(E_Q-E_R)\tau_2},
\end{align}
where $ \hat{Z}_{\mu_i}=\hat{\sigma}^{z}_{\mu_i ,\text{odd}}+\hat{\sigma}^{z}_{\mu_i,\text{even}} $, is the sum of Pauli matrices of the two nearest neighbor spins on z-link of the $\mu$ unit cell, see Fig.~\ref{Majorana_paring}. The matrix elements in Eq.(\ref{R_12_general}) are identical for all correlation functions, with the only difference being their time dependencies\cite{choi1,Negahdari_Langari_2023}. For example, in functions $\Upsilon^{(3),z}_{zzz}(\tau_2,\tau_1,0) $ and
$\Upsilon^{(3),z}_{zzz}(\tau_2,0,\tau_1) $, the time dependencies are
$e^{i(E_Q-E_G)\tau_1+i(E_Q-E_R)\tau_2} $ and
$ e^{i(E_P-E_G)\tau_1+i(E_Q-E_R)\tau_2}$,
respectively.

Our calculations to obtain the nonlinear susceptibilities in 2DCS are done within the few-matter fermion approximation, which has been
justified in Sec.\ref{linear_response_sec}. Given that the ground state in Eq.(\ref{R_12_general}) has no excitations, the states $|P\rangle$ and $|R\rangle$, as well as state $|Q\rangle$, have odd and even parity $\pi_a$, respectively. According to the few-matter fermion approximation, it suffices to consider the first-order excitations for these states and neglect the higher-order ones; specifically, this means considering single-matter fermion excitations for states $|P\rangle$ and $|R\rangle$, and zero-matter fermion excitations for state $|Q\rangle$. Similar calculations has been done in references [\onlinecite{choi1}] and [\onlinecite{Negahdari_Langari_2023}].

Fourier transform of the third-order susceptibilities at the isotropic coupling for three distinct  models are plotted in Fig.~\ref{nonlinear_response}, namely: the Kekulé-Kitaev model and the model in the presence of weak uniform and staggered magnetic fields (see appendix \ref{Matrix_elements} for details of computations). 
Our computations for the Kekulé-Kitaev model have been performed with geometric parameters 
$L_1 = L_2 = 34, M=0$, and in the presence of magnetic fields we adopted $L_1 = L_2 = 30, M=0$. These system sizes are sufficiently large to reveal the expected features of the non-linear response correctly. 

\begin{figure}[!h]
	\includegraphics[width=0.6\columnwidth]{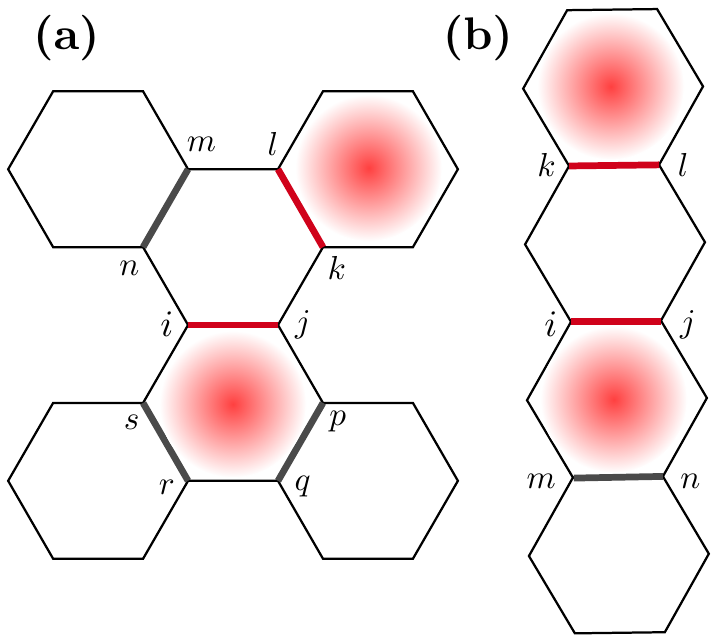}
	\caption{In the standard gauge configuration ($u_{ij}=+1$), two non-adjacent fluxes can be created by flipping two  $u_{ij}$'s. In the Kekulé-Kitaev model, panel (a), there exist four such states per unit cell. One of these states is created by setting  $u_{ij}=-1$ and $u_{kl}=-1$, as depicted in this figure. The remaining three states arise from flipping pairs $(u_{ij},u_{mn})$, $(u_{ij},u_{pq})$, and $(u_{ij},u_{rs})$. In contrast, the Kitaev model, panel (b),  exhibits two such states per each unit cell.}
	\label{KekuleKitaevFlux}
\end{figure}
 In the nonlinear response of the Kitaev model, the streak signals serve as signatures of the itinerant Majorana fermions, while their intercepts correspond to the localized ones\cite{choi1}. Similarly, such signals indicated by gray lines in Figs.~\ref{nonlinear_response}(d,e), are also present in the Kekulé-Kitaev model and the model in the presence of a weak uniform magnetic field.  The streak signals located in the second and fourth quadrants of Figs.~\ref{nonlinear_response}(d,e)  originate from the functions 
 $\Upsilon^{(2),z}_{zzz}(\omega_2,0,\omega_1)=\Upsilon^{(3),z}_{zzz}(\omega_2,0,\omega_1)$. These functions possess the following characteristic frequencies,
 \begin{equation}
 	\label{R23Freq}
 	\begin{aligned}
 		 	\omega_1=E_G-E_P=E_0-E_2-\varepsilon_p ,\\
 		  	\omega_2=E_R-E_Q=E_2-E_4+\varepsilon_r.\\
 	\end{aligned}
 \end{equation}
 According to reference [\onlinecite{choi1}] and based on Eq.(\ref{R23Freq}), when $|P\rangle=|R\rangle$, constructive interference occurs, which generates the streak signal $\omega_1+\omega_2=-(E_4-E_0)$.  In a weak uniform magnetic field, Fig.~\ref{nonlinear_response}(e), the streak signals are weaker than those in the original model, Fig.~\ref{nonlinear_response}(d). According to Fig.~\ref{stacture_factor_q0}(b), this behavior may be explained with the observation that only the first few single-matter fermion excitations have dominant contributions, whereas higher energy single-matter fermion  excitations exhibit less impacts.

 All of $\Upsilon^{(2),z}_{zzz}(\omega_2,\omega_1,0) $ functions  in $\chi^{(3),z}_{zzz}(\omega_2,\omega_1,0)$ contribute to the generation of vertical signals in Fig.~\ref{nonlinear_response}(a,b). For example, the characteristic frequencies in function 
 	$ \Upsilon^{(3),z}_{zzz}(\omega_2,\omega_1,0)$
 are as follows,
 \begin{equation}
 	\label{R3tt0Fun}
 	\begin{aligned}
 		 	&\omega_1=E_G-E_Q=E_0-E_4 ,\\
 		 	&\omega_2=E_R-E_Q=E_2-E_4+\varepsilon_r.
 	\end{aligned}
 \end{equation}
As indicated in Eq.(\ref{R3tt0Fun}), $\omega_1$ is a fixed value, while $\omega_2$ takes on various values due to the presence of $\varepsilon_r$, resulting in the generation of a vertical signal. Furthermore, part of the horizontal and vertical broadening observed in the signals of Fig.~\ref{nonlinear_response} arises from the one-sided Fourier transform, which occurs because $\tau_1,\tau_2>0$. For further clarification on this topic,  see reference [\onlinecite{Nandkishore_Choi2021}].

The nonlinear responses in the presence of a weak staggered field are shown in Figs.~\ref{nonlinear_response}(c,f).  As evident from these figures, characteristic streak signals are no longer present; instead, distinct sharp peaks emerge, which arise from the Majorana bound states and consequently indicate the presence of non-Abelian anyons\cite{Negahdari_Langari_2023}. The reason for absence of the streak signals is that the few eigenstates with energies inside the fermionic gap, as illustrated in Fig.~\ref{bound_states}(b), or in other words, the Majorana bound states, have significantly lower energies compared to the states above the gap. Consequently, the corresponding matrix elements for these states are substantially stronger than those for other states present in the continuum of excitations above the gap\cite{Negahdari_Langari_2023}.
	
The nonlinear responses in Figs.~\ref{nonlinear_response}(c,f) is nearly identical to that of the Kitaev model in the non-Abelian phase\cite{Negahdari_Langari_2023}. Each of the distinct signals observed in Fig.~\ref{nonlinear_response}(c,f) arises from different flux configurations of non-Abelian anyons. For example, the signals marked with green circles in these two figures arise from the  flux configurations with  two non-adjacent fluxes. One of these flux configurations are depicted in Fig.~\ref{KekuleKitaevFlux}(a). If we compare the  nonlinear responses reported for the Kitaev model in reference [\onlinecite{Negahdari_Langari_2023}]  with the nonlinear responses shown in Fig.~\ref{nonlinear_response}(c,f), we find that the strength of the signals arising from the two non-adjacent fluxes in the Kitaev model is comparable to that of other signals. In contrast, these signals (marked with green circles) in the Kekulé-Kitaev model are more pronounced than   the other signals. This discrepancy, as illustrated in Fig.~\ref{KekuleKitaevFlux}, arises from the fact that the density of such flux configurations per unit cell in the Kekulé-Kitaev model is larger than their density in the Kitaev model.


\section{Discussions and Conclusions}\label{conclusion}

In this work, we studied a variant of the Kitaev model known as the Kekulé-Kitaev model. Initially, we derived an explicit relation for identifying physical and unphysical states within the framework of Majorana fermions representation. By calculating the spectrum and energy bands, we
 found that a weak uniform magnetic field, as a simple perturbation, does not create a gap in the Kekulé-Kitaev model's spectrum but instead lifts the degeneracy of the two gapless Dirac cones. In contrast, a weak staggered magnetic field breaks the gapless nature of these Dirac cones, opening a gap. The stability of gapless Dirac points has been examined, which is preserved if the combination of time reversal symmetry ($\hat{\tau}$) and space inversion ($\hat{I}$) symmetry is not broken upon the introduction of a perturbation, i.e., $\hat{\tau}\hat{I}$.  The space inversion is the product of inversion and a global spin rotation by the angle 
$\pi/2$ around any Cartesian axis.
	
In the presence of a staggered magnetic field, flux excitations create Majorana bound states within the gap of the Kekulé-Kitaev model. These bound states are responsible for the non-Abelian statistics of the flux excitations. This behavior contrasts with the Kitaev model, which hosts the  non-Abelian phase under the uniform magnetic field. In fact, the response of the Kekulé-Kitaev and Kitaev, in uniform and staggered magnetic fields are opposites to each other. Therefore, uniform, staggered, or other types of magnetic fields act as simple perturbations that can induce non-Abelian phases in various generalizations of the Kitaev model.	
	
In the second half of the work, we investigated the linear and non-linear responses of the Kekulé-Kitaev model by calculating the dynamical spin structure factor and two-dimensional coherent spectroscopy.
Due to the lack of long-range order and the presence of fractional excitations, the dynamical spin structure factor exhibits a broad spectrum and diffusive pattern. This linear response is nearly identical to that of the Kitaev model.
 In the non-linear response, matter and gauge excitations show distinct signatures, visible as streak signals and their intercepts, respectively. We used the single-matter fermion excitation approximation to calculate both linear and non-linear responses. To verify its validity, we assessed the contribution of three-matter fermion excitations in the dynamical spin structure factor. Our results show that the contributions from many-matter  fermion excitations are  are negligible compared to those from single-matter fermion ones, confirming the validity of the single-matter fermion excitation approximation.

Given the similarities between the Kekulé-Kitaev and Kitaev models, a natural question is whether the Kekulé-Kitaev model can be transformed into the Kitaev model through a suitable unitary transformation? We demonstrated that the ground state of the Kekulé-Kitaev model at isotropic coupling, via  unitary spin rotations, transforms into the excited state of the Kitaev model with a uniform flux density $\rho_f = 2/3$.
The unitary transformation defines the relation between the spectrum of the two models. The detail of this transformation is given in appendix \ref{mapping}. The bulk properties of the Kekulé-Kitaev model are expected to fall within the Kitaev’s sixteenfold way as the Kitaev model\cite{Kitaev_2006,Zhang2020}. Because, both models are described by free Majorana fermions in the background of $Z_2$ gauge fields on a honeycomb lattice. This expectation is further supported by the fact that the model at isotropic coupling, transforms into the Kitaev model via a unitary transformation.

As shown in Fig.~\ref{energy_bands}(b), the low-energy physics of the  Kekulé-Kitaev model in the uniform field is no longer described by Dirac crossings. Therefore, an important question arises: what is the role of these non-Dirac crossings in the low-energy properties of the system? Furthermore, our study did not take into account the effects of disorder and finite temperature, and further research could be conducted to explore their impact.

\section*{ Acknowledgments}
The authors would like to acknowledge M. Kargarian and  A. Vaezi  for  fruitful comments and discussions.

\newpage
\appendix
\begin{widetext}

\section{Label of sites}
\label{SiteNoPhase}

For a lattice with $N$ unit cells, there are $6N$ sites, which takes the labels from $1$ to $6N$. According to our notation, Table~\ref{table}  gives the labels and coordinates of all sites. In this table, $j$ is the label of a unit cell, which takes the values from $0$ to $N-1$, and  $\boldsymbol{R}_i$ points to the center of a unit cell, where  site $i$ belongs to that,
\begin{equation}
	\label{}
	\boldsymbol{R_i}=\bigg(\bigg[\frac{i-1}{6}\bigg]\ \text{mod}\ L_1\bigg)\ \boldsymbol{a}_1+\bigg[\frac{(i-1)/6}{L_1}\bigg]\ \boldsymbol{a}_2,
\end{equation}
where $ [\dots] $ denotes the integer-valued function.
 \begin{table}[!h]
	\caption{} 
	\centering 
	\begin{tabular}{c c c} 
		\hline\hline 
		sublattice &
		label &
		position 
		\\[1ex]
		\hline 
		A&$ 6j+1 $ &$ \boldsymbol{R}_i+\boldsymbol{e}_1 $ \\
		B&$ 6j+2 $&$ \boldsymbol{R}_i-\boldsymbol{e}_3 $ \\
		C&$ 6j+3 $ &$ \boldsymbol{R}_i+\boldsymbol{e}_2$ \\
		D&$6j+4$&$\boldsymbol{R}_i-\boldsymbol{e}_1 $\\
		E&$ 6j+5 $&$ \boldsymbol{R}_i+\boldsymbol{e}_3 $\\
		F& $6j+6 $&$ \boldsymbol{R}_i-\boldsymbol{e}_2$\\
		\hline 
	\end{tabular}
	\label{table} 
\end{table} 


\section{Diagonalization in the momentum space}
\label{energyBands}
The eigenvalues of Hamiltonian (\ref{momentumHam}) are obtained by the following equation
\begin{equation}
	\label{pure_det}
	\det\begin{pmatrix}
		-E_{\boldsymbol{k}}\mathds{1}&  iM_{\boldsymbol{k}}\\
		-iM_{\boldsymbol{k}}^{\dagger}& -E_{\boldsymbol{k}}\mathds{1}
	\end{pmatrix}=0.
\end{equation}
The above equation is simplified using a determinant identity as follows,
\begin{equation}
	\label{pure_det_simple}
	\det\bigg[E_{\boldsymbol{k}}^2\mathds{1}-M_{\boldsymbol{k}}^{\dagger}M_{\boldsymbol{k}}^{}\bigg]=0.
\end{equation} 
For brevity, we express the matrix 
$ M_{\boldsymbol{k}}^{\dagger}M_{\boldsymbol{k}}^{} $ 
as,
 \begin{equation}
 	\begin{aligned}
 		M_{\boldsymbol{k}}^{\dagger}M_{\boldsymbol{k}}^{}=
 		\begin{pmatrix}
 			J_0^2&A^{}_{\boldsymbol{k}}&B^{}_{\boldsymbol{k}}\\
 			A^{*}_{\boldsymbol{k}}&J_0^2&C^{}_{\boldsymbol{k}}\\
 			B^{*}_{\boldsymbol{k}}&C^{*}_{\boldsymbol{k}}&J_0^2
 		\end{pmatrix} 
 	\end{aligned} \;\; ,\qquad\text{with}\quad
 J_0^2=4(J_x^2+J_y^2+J_z^2).
  \end{equation}
Hence,  Eq.(\ref{pure_det_simple})  leads to the following cubic equation,	
\begin{equation}
	\label{pure_cubic}
	x^3_{\boldsymbol{k}}+p_{\boldsymbol{k}}^{}x_{\boldsymbol{k}}^{}+q_{\boldsymbol{k}}=0,\quad x_{\boldsymbol{k}}=E^2_{\boldsymbol{k}}-J_0^2,
\end{equation}
with,
\begin{equation}
	\begin{aligned}
		&p_{\boldsymbol{k}}=-(|A_{\boldsymbol{k}}|^2+|B_{\boldsymbol{k}}|^2+|C_{\boldsymbol{k}}|^2),\\
		&q_{\boldsymbol{k}}=-(A_{\boldsymbol{k}}^{} B_{\boldsymbol{k}}^{*}C^{}_{\boldsymbol{k}}+A_{\boldsymbol{k}}^{*}B_{\boldsymbol{k}}^{}C_{\boldsymbol{k}}^{*} ).	
	\end{aligned}
\end{equation}
The cubic equation (\ref{pure_cubic}) has three roots, which gives six energy bands,
\begin{equation}
	\begin{aligned}
		&x_{n,\boldsymbol{k}}=2\sqrt{\frac{-p_{\boldsymbol{k}}}{3}}\cos\bigg[\frac{1}{3}\cos^{-1}\bigg(\frac{3q_{\boldsymbol{k}}}{2p_{\boldsymbol{k}}\sqrt{-p_{\boldsymbol{k}}/3}}\bigg)+n\frac{2\pi}{3}\bigg] ,\\
		&E_{n,\boldsymbol{k}}=\pm\sqrt{x_{n,\boldsymbol{k}}+J_0^2} \;\; ,\quad n=1,2,3.
	\end{aligned}
\end{equation}


\section{Calculation of $ \hat{D} $ operator} \label{D_operator}
We generalize the approach given in [\onlinecite{Pedrocchi_Loss}] to the Kekulé-Kitaev model to obtain $\hat{D}$. This operator is given as follows,
\begin{equation}
	\hat{D}=\prod_{i=1}^{6N}\hat{b}_i^{x}\hat{b}_i^{y}\hat{b}_i^{z}\hat{c}_i=\prod_{i=1}^{6N}\hat{b}_i^{x}\prod_{i=1}^{6N}\hat{b}_i^{y}\prod_{i=1}^{6N}\hat{b}_i^{z}\prod_{i=1}^{6N}\hat{c}_i .
\end{equation}
Similar steps as given in Ref.[\onlinecite{Pedrocchi_Loss}], leads to the following  equation,
\begin{equation}
	\prod_{i=1}^{6N}\hat{c}_i=(i)^{3N}\det{(\Lambda_u)}\hat{\pi}_a,
\end{equation}
where $\hat{\pi}_a=\prod_{k=1}^{3N}(1-2\hat{a}^{\dagger}_k\hat{a}_k^{})$ is the parity of the matter fermions. To derive a simple relation for the remaining factors, we need to pair up $\hat{b}_i^{\alpha}$ operators, which are related to the same $\alpha$-bond. By doing this, $ \prod_{i=1}^{6N}\hat{b}_i^{\alpha} $  is reduced to the product of $ \hat{u}_{\langle jk\rangle}^{\alpha} $. Given that the y- and z-links are placed in the unit cells, it is straightforward to derive the following relation,
\begin{equation}
	\prod_{i=1}^{6N}\hat{b}^{\alpha}_i=(-i)^{3N}\hat{\pi}_{\chi_{\alpha}}\quad,\quad \alpha=y,z.
\end{equation}
$\hat{\pi}_{\chi_{\alpha}}=\prod_{\langle jk\rangle_{\alpha}}^{}\hat{u}_{\langle jk\rangle}^{\alpha}$ is the parity of the complex gauge fermions. According to our convention, the first index in  $\hat{u}_{\langle jk\rangle}^{\alpha}$ belongs to odd sites, and the second one belongs to even sites. Given that the x-links are located outside the unit cells, the pairing of the $ \hat{b}^x_i $ operators in the expression $ \prod_{i=1}^{6N}\hat{b}^{x}_i $ depends on the boundary conditions. To simplify the pairing procedure, we decompose the expression into $ L_2 $  products,
\begin{equation}
	\prod_{i=1}^{6N}\hat{b}^{x}_i=\prod_{n=1}^{L_2}\hat{\mathcal{T}}_n\quad,\quad
	\hat{\mathcal{T}}_n=\prod_{j=1}^{6L_1}\hat{b}_{6L_1(n-1)+j}^{x}.
\end{equation}
$ \hat{\mathcal{T}}_n $  is the product of $ 6L_1 $  operators existing in each of the $ n$th-row of the lattice, which is schematically depicted in Fig~\ref{D_operator_two_flux_z_and_xy}(a). Some of $ \hat{b}^{x}_i $'s are paired  within each row, while some of them are paired between two successive rows  $ \{n,n+1; n\neq 1,L_2\} $, and on boundaries $ \{1,L_2\} $, giving rise to the three distinct phases $ (-1)^{L_2(L_1-1)}$,   $ (-1)^{L_2-1} $, and  $ (-1)^{L_1-M} $, respectively. So,
\begin{equation}
	\prod_{i=1}^{6N}\hat{b}^{x}_i=(-i)^{3N}(-1)^{N}(-1)^{L_1-M-1}\hat{\pi}_{\chi_{x}}.
\end{equation}
Finally, we obtain the following relation,
\begin{equation}
\hat{D}=(-1)^{\theta}\det(\Lambda_u)\hat{\pi}_{\chi}\hat{\pi}_a,
\end{equation}
where $ \theta=L_1-M-1 $ and $ \hat{\pi}_{\chi}=\hat{\pi}_{\chi_x}\hat{\pi}_{\chi_y}\hat{\pi}_{\chi_z}$.

\begin{figure*}[!t]
	\centering
	\includegraphics[scale=0.93]{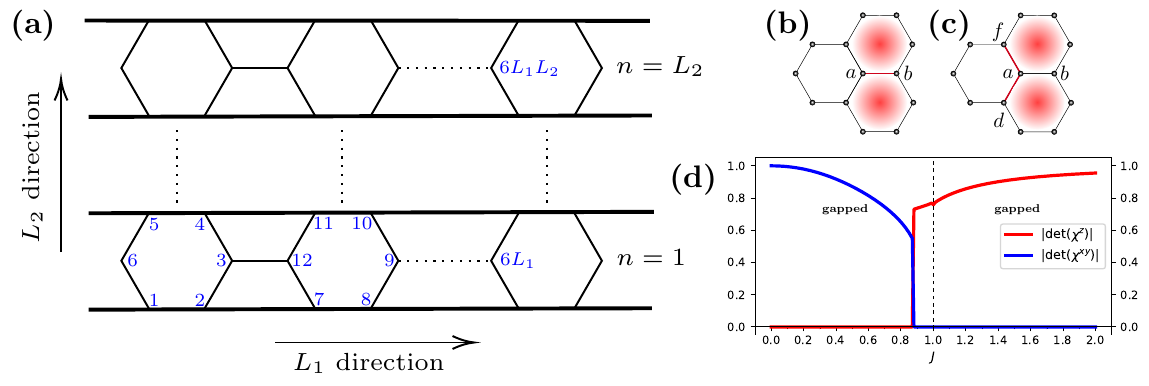}
	\caption{(a) The lattice can be divided into $L_2$ rows from $1$ to $L_2$. Two different configurations, (b)  $|P^{z}_{\mu_1}\rangle$ and (c) $|P^{xy}_{\mu_1}\rangle$, for two-adjacent fluxes. (d) Dynamical phase diagram of the original  Kekulé-Kitaev model for a system with $L_1=40$, $L_2=41$, and $M=14$. The red line represents the overlap of the vacuum state of $|P^{z}_{\mu_1}\rangle$, while the blue line represents the overlap of the vacuum state of $|P^{xy}_{\mu_1}\rangle$  with the ground state $|G\rangle$.}
	\label{D_operator_two_flux_z_and_xy}
\end{figure*}


\section{Examination of $\hat{\tau}\hat{I}$ symmetry}\label{StabilityOfDirac}
 To examine the existence of $\hat{\tau}\hat{I}$ symmetry, we expand the Hamiltonians close to the $\Gamma$  point, where the Dirac points are located.  Specifically, in the vicinity of this point, the Hamiltonian is given by
 \begin{equation}
 	\label{Ham_eff}
 	\tilde{H}^{\{u\}}_{\text{eff}}(\boldsymbol{k})=\tilde{H}^{\{u\}}_{\text{eff},1}(\boldsymbol{k})+\tilde{H}^{\{u\}}_{\text{eff},2}(\boldsymbol{k}),\quad \text{with}\quad\tilde{H}^{\{u\}}_{\text{eff},1}(\boldsymbol{k})=
 	i\begin{pmatrix}
 		0&M(\boldsymbol{k})\\
 		-M^{\dagger}(\boldsymbol{k})&0
 	\end{pmatrix},\quad
 	\tilde{H}^{\{u\}}_{\text{eff},2}(\boldsymbol{k})=
 	i\begin{pmatrix}
 		F_o(\boldsymbol{k})&0\\
 		0&F_e(\boldsymbol{k})
 	\end{pmatrix},
 \end{equation}
where $\tilde{H}^{\{u\}}_{\text{eff},1}(\boldsymbol{k})$ represents the first-neighbor hopping and $\tilde{H}^{\{u\}}_{\text{eff},2}(\boldsymbol{k})$ the second-neighbor hopping. $M(\boldsymbol{k}) $, $F_o(\boldsymbol{k}) $, and $F_e(\boldsymbol{k}) $ are given by the following matrices,
\begin{equation}
M(\boldsymbol{k})=\begin{pmatrix}
		J_z&J_x(1-  \frac{i}{2}k_x -  \frac{i\sqrt{3}}{2}k_y )&J_y\\
		J_y&J_z&J_x(1+ik_x)\\
	    J_x(1-\frac{i}{2}k_x +  \frac{i\sqrt{3}}{2}k_y )&J_y&J_z
	\end{pmatrix},
\end{equation}
\begin{equation}
	F_o(\boldsymbol{k})=
\begin{pmatrix}
	0&K(3-  \frac{3i}{2}k_x -  \frac{i\sqrt{3}}{2}k_y )&K(-3+i  \sqrt{3}k_y)\\
	K(-3-  \frac{3i}{2}k_x - \frac{i\sqrt{3}}{2}k_y )&0&
	K(3+\frac{3i}{2}k_x - \frac{i\sqrt{3}}{2}k_y )\\
	K(3+i\sqrt{3}k_y)&K(-3+\frac{3i}{2}k_x -  \frac{i\sqrt{3}}{2}k_y )&0
\end{pmatrix},
\end{equation}
\begin{equation}
F_e(\boldsymbol{k})=	\begin{pmatrix}
	0&K(\pm 3\mp i  \sqrt{3}k_y)&K(\mp3\mp  \frac{3i}{2}k_x \pm  \frac{i\sqrt{3}}{2}k_y )\\
	K(\mp3\mp i\sqrt{3}k_y)&0&K(\pm3\pm \frac{3i}{2}k_x \pm  \frac{i\sqrt{3}}{2}k_y )\\
	K(\pm3\mp\frac{3i}{2}k_x\pm\frac{i\sqrt{3}}{2}k_y )&K(\mp3\pm  \frac{3i}{2}k_x \pm  \frac{i\sqrt{3}}{2}k_y )&0
\end{pmatrix},
\end{equation}
where, in $F_e(\boldsymbol{k}) $, the upper and lower signs correspond to uniform and staggered magnetic fields, respectively.
In the original Kekulé-Kitaev model,  $K=0$ or equivalently
$\tilde{H}^{\{u\}}_{\text{eff},2}(\boldsymbol{k})=0$. As mentioned in Sec.\ref{Stability}, the $\hat{I}$ symmetry is a simultaneous combination of
 inversion (or a $\pi$ rotation of the lattice around the center of any hexagonal plaquette $W_{\alpha}$) and a global spin rotation by the angle
 $\pi/2$ around the $\alpha$-axis, where $\alpha=x,y,z$.
Under the operation of $\hat{I}$, the Majorana fermion operators transform as
\begin{equation}
	\hat{c}_{\mu,A}\leftrightarrow \hat{c}_{\nu,D},\quad
	\hat{c}_{\mu,C}\leftrightarrow \hat{c}_{\nu,F},\quad
	\hat{c}_{\mu,E}\leftrightarrow \hat{c}_{\nu,B}.
\end{equation}
This basis transformation can be represented by the following matrix,
\begin{equation}
	I_m=\begin{pmatrix}
		0&0&0&0&1&0\\
		0&0&0&0&0&1\\
		0&0&0&1&0&0\\
		0&0&1&0&0&0\\
		1&0&0&0&0&0\\
		0&1&0&0&0&0
	\end{pmatrix}.
\end{equation}
Therefore, if Hamiltonian (\ref{Ham_eff}) has
$ \hat{\tau}\hat{I}$-symmetry, where $\tau$ is time-reversal opertation, the following relation holds
\begin{equation}
	\label{TimeInversiosn}
	(\hat{\tau}\hat{I}) \tilde{H}^{\{u\}}_{\text{eff}}(\boldsymbol{k}) (\hat{\tau}\hat{I})^{-1}=\tilde{H}^{\{u\}}_{\text{eff}}(\boldsymbol{k})\quad
	\rightarrow\quad
	I_m^T (\tilde{H}^{\{u\}}_{\text{eff}}(\boldsymbol{k}))^*I_m=\tilde{H}^{\{u\}}_{\text{eff}}(\boldsymbol{k}),
\end{equation}
where in the right identity, $T$ and  $*$ represent the transpose and complex conjugate operations, respectively. We can easily verify that in the original Kekulé-Kitaev model, and the model in the presence of uniform magnetic field, equality (\ref{TimeInversiosn}) holds. However, in the presence of the staggered magnetic field the Hamiltonian is not invariant upon $ \hat{\tau}\hat{I}$ operation.

Given that under the $\hat{I}$ transformation, the odd sublattices transform into even ones and vice versa,  the transformation $u_{ij}\rightarrow -u_{ij}$ must be applied to the gauge degrees of freedom. On the other hand, under time-reversal transformation ($\hat{\tau}$),  $u_{ij}\rightarrow -u_{ij}$ also occurs. Therefore, no extra negative sign in Eq.(\ref{TimeInversiosn}) is generated by the $\hat{\tau}\hat{I}$ transformation. It is worth noting that under time-reversal, the Majorana fermion operators transform as follows
\begin{equation}
	\hat{\tau}\hat{c}_j\hat{\tau}^{-1}=-\hat{b}^y_j,\quad
	\hat{\tau}\hat{b}_j^y\hat{\tau}^{-1}=-\hat{c}_j,\quad
	\hat{\tau}\hat{b}_j^x\hat{\tau}^{-1}=\hat{b}^z_j,\quad
	\hat{\tau}\hat{b}_j^z\hat{\tau}^{-1}=\hat{b}^x_j.
\end{equation}

\section{Matrix elements in linear and non-linear responses }\label{Matrix_elements}
The calculation of the matrix element $ \langle P|\hat{\sigma}_a|G\rangle $ for both even and odd parity of matter fermions in $ |P\rangle $ state can be addressed by considering the gauge redundancy. Suppose $ a\in A $ sublattice and according to the gauge structure of the Hamiltonian, for $ |P\rangle $ state,  it is possible to choose two initial gauge configurations—one characterized by an odd parity of gauge fermions and the other by an even parity. These configurations are depicted in Figs.~\ref{D_operator_two_flux_z_and_xy}(b,c),
\begin{equation}
	\begin{aligned}
		&|p_{\mu_1}^z\rangle=\hat{\chi}^{\dagger}_{\langle ab\rangle_z}|\mathcal{G}\rangle|M_{\mu_1}^z\rangle\\
		&|p_{\mu_1}^{xy}\rangle=\hat{\chi}^{\dagger}_{\langle ad\rangle_x}\hat{\chi}^{\dagger}_{\langle af\rangle_y}|\mathcal{G}\rangle|M_{\mu_1}^{xy}\rangle.
	\end{aligned}
\end{equation}
In order to construct the physical states, we have to sum over all gauge configurations, i.e. $|P_{\mu_1}^z\rangle=\hat{\mathcal{S}}|p_{\mu_1}^{z}\rangle$ and $|P_{\mu_1}^{xy}\rangle=\hat{\mathcal{S}}|p_{\mu_1}^{xy}\rangle$, which is given by
\begin{equation}
	\begin{aligned}
		&|P_{\mu_1}^z\rangle=\bigg[\hat{\chi}^{\dagger}_{\langle ab\rangle_z}- \hat{\chi}_{\langle ad\rangle_x}^{\dagger}\hat{\chi}_{\langle af\rangle_y}^{\dagger} \hat{c}_a+\dots\bigg]|\mathcal{G}\rangle|M_{\mu_1}^z\rangle,\\
		&|P_{\mu_1}^{xy}\rangle=\bigg[
		\hat{\chi}^{\dagger}_{\langle ad\rangle_x}\hat{\chi}^{\dagger}_{\langle af\rangle_y}-\hat{\chi}_{\langle ab\rangle_z}^{\dagger}\ \hat{c}_a+\dots\bigg] |\mathcal{G}\rangle|M_{\mu_1}^{xy}\rangle.
	\end{aligned}
\end{equation}
Utilizing straightforward calculations, we derive the subsequent relations,
\begin{equation}
	\label{different_mat_elements}
	\begin{aligned}
		&\langle P_{\mu_1}^z|\hat{\sigma}_a^z|G\rangle=i\langle M_{\mu_1}^{z}|\hat{c}_a|\mathcal{M}_0\rangle,\\
		&\langle
		P_{\mu_1}^z|\hat{\sigma}_b^z|G\rangle=\langle  M_{\mu_1}^z|\hat{c}_b|\mathcal{M}_0\rangle,\\
		&\langle P_{\mu_1}^{xy}|\hat{\sigma}_a^z|G\rangle=
		-i\langle M_{\mu_1}^{xy}|\mathcal{M}_0\rangle,\\
		&\langle P_{\mu_1}^{xy}|\hat{\sigma}_b^z|G\rangle=
		-\langle M_{\mu_1}^{xy}|\hat{c}_a\hat{c}_b|\mathcal{M}_0\rangle.
	\end{aligned}
\end{equation}
According to Eq.(\ref{different_mat_elements}), when the parity of matter fermions in $ |P\rangle $ state is odd, choosing  $  |P_{\mu_1}^z\rangle$ representation is necessary to obtain a non-zero matrix element. Conversely, if the parity is even, $ |P_{\mu_1}^{xy}\rangle $ representation must be chosen. Fig.~\ref{D_operator_two_flux_z_and_xy}(d)  shows the dynamical phase diagram of the Kekulé-Kitaev model.  The parity in $ |P\rangle $ state remains even up to $J=0.87$, and changes to odd parity thereafter, where we set  $J_z=1$ and $J=\frac{J_x}{J_z}=\frac{J_y}{J_z}$. For a similar discussion, see Refs. [\onlinecite{Knolle_Chalker_Moessner2015,Knolle2016_theses}].

In the context of single-matter fermion excitations, the matter state is represented by $ |M_{\mu_1}^z\rangle=\hat{\overline{a}}^{\dagger}_r |\mathcal{M}_{\mu_1}^z\rangle $. The calculation of matrix elements within the single-matter fermion approximation for the Kekulé-Kitaev model  is similar to the detailed discussions provided in Refs. [\onlinecite{Knolle_Chalker_Moessner2015,Knolle2016_theses}] for the Kitaev model. To keep it concise, we only show the final outcome. For example, in the Kekulé-Kitaev model,
\begin{equation}
	\begin{aligned}
		&\langle P_{\mu_1}^z|\hat{\sigma}_{a}^z|G\rangle=
		i\sqrt{|\det(\mathcal{X})|}\ 
		\bigg[
		U_0\mathcal{X}^{-1}
		\bigg]_{\mu_a r},\\
		&\langle R_{\mu_i}^z|\hat{\sigma}_{a}^z|Q_{\mu_1\mu_i}^{zz}\rangle=
		i\sqrt{|\det(\mathcal{X}')|}\ 
		\bigg[
		U_4
		\mathcal{X}'^{-1}
		\bigg]_{\mu_a r},
	\end{aligned}
\end{equation}
where $\mathcal{X}=\frac{1}{2}(U^T_2U_0^{}+V^T_2V_0^{})$ and 
$\mathcal{X}'=\frac{1}{2}(U^T_2U_4^{}+V^T_2V_4^{})$\cite{choi1}. Moreover, $ \mu_a=\mu $ and index $i$ in $ \mu_i$ takes the values of $1,2,3$; see Fig.~\ref{Majorana_paring}. It should be noticed that  $ \mu_c=\mu +N$ and $ \mu_e=\mu +2N$.  In the above relations, if $\hat{\sigma}_a^z \rightarrow \hat{\sigma}_b^z  $, we perform the substitutions $ U_0\rightarrow V_0 $ and $ U_4\rightarrow -V_4 $.  Similarly, we can compute  these matrix elements in the presence of weak uniform and staggered magnetic fields. More dtails is available in Ref.[\onlinecite{Negahdari_Langari_2023}].

In order to examine the validity of the single-matter fermion  approximation, we compute the contribution of three-matter fermion excitations in the structure factor. Specifically, within the context of three-matter fermion excitations, the matter  part of $ |P_{\mu_1}^z\rangle $  takes the form $ |M_{\mu_1}^z\rangle=\hat{\overline{a}}^{\dagger}_r\hat{\overline{a}}^{\dagger}_s\hat{\overline{a}}^{\dagger}_q |\mathcal{M}_{\mu_1}^z\rangle $.  We summarize the matrix elements pertaining to these excitations as follows,
\begin{equation}
	\label{three_matter_mat_element}
		\langle P_{\mu_1}^z|\hat{\sigma}_{a}^z|G\rangle=i\sqrt{|\det(\mathcal{X})|}\ 
		\bigg[(U_0\mathcal{X}^{-1})_{\mu_ar}(\mathcal{Y}^{}\mathcal{X}^{-1})_{sq}
		-(U_0\mathcal{X}^{-1})_{\mu_as}(\mathcal{Y}^{}\mathcal{X}^{-1})_{rq}
		+U_0\mathcal{X}^{-1})_{\mu_aq}(\mathcal{Y}^{}\mathcal{X}^{-1})_{rs}\bigg].
\end{equation}
For the matrix element of B-sublattice ($ \hat{\sigma}_{a}^z\rightarrow \hat{\sigma}_{b}^z$), we perform the substitution  $ U_0\rightarrow V_0$. The straightforward result in Eq.(\ref{three_matter_mat_element}) emerges as the sum of 40 terms, which we categorize into three distinct parts,
\begin{equation}
	\begin{aligned}
		&\text{(i) : }\langle \hat{a}_{q'}^{}\hat{a}_{s'}^{}\hat{a}_{r'}^{\dagger}\hat{a}_{r''}^{\dagger}\rangle,
		\langle \hat{a}_{q'}^{}\hat{a}_{s'}^{\dagger}\hat{a}_{r'}^{}\hat{a}_{r''}^{\dagger}\rangle,
		\langle \hat{a}_{q'}^{\dagger}\hat{a}_{s'}^{}\hat{a}_{r'}^{}\hat{a}_{r''}^{\dagger}\rangle,\\
		&\text{(ii) : }\langle \hat{a}_{m}^{}\hat{a}_{n}^{}\hat{a}_{q'}^{}\hat{a}_{s'}^{\dagger}\hat{a}_{r'}^{\dagger}\hat{a}_{r''}^{\dagger}\rangle,
		\langle \hat{a}_{m}^{}\hat{a}_{n}^{}\hat{a}_{q'}^{\dagger}\hat{a}_{s'}^{}\hat{a}_{r'}^{\dagger}\hat{a}_{r''}^{\dagger}\rangle,
	 \langle\hat{a}_{m}^{}\hat{a}_{n}^{}\hat{a}_{q'}^{\dagger}\hat{a}_{s'}^{\dagger}\hat{a}_{r'}^{}\hat{a}_{r''}^{\dagger}\rangle,\\
		&\text{(iii) : }\langle \hat{a}_{m}^{}\hat{a}_{n}^{}\hat{a}_{q'}^{}\hat{a}_{s'}^{\dagger}\hat{a}_{r'}^{\dagger}\hat{a}_{r''}^{\dagger}\rangle.
	\end{aligned}
\end{equation}
Fig.~\ref{stacture_factor_q0}(c) shows the components of the structure factor in the gapped phase. For the calculation of  $ S^{zz}(0,\omega) $, it is crucial to note that states  $ |G\rangle $ and $ |P\rangle $ have the same matter fermion parity. Consequently, the two-flux state must necessarily include an even number of matter excitations. As previously mentioned, adopting $ |P^{xy}_{\mu_1}\rangle $ representation is essential to yield a non-zero matrix element. Specifically, for the case of zero-matter fermion  excitation, i.e., $ |M_{\mu_1}^{xy}\rangle=|\mathcal{M}_{\mu_1}^{xy}\rangle $,
\begin{equation}
	\label{ZeroMatter}
	\begin{aligned}
		&\langle P_{\mu_1}^{xy}|\hat{\sigma}_{a}^z|G\rangle=
		-i\sqrt{|\det(\mathcal{X})|},\\
		&\langle P_{\mu_1}^{xy}|\hat{\sigma}_{b}^z|G\rangle=
		-i\sqrt{|\det(\mathcal{X})|}
		\ 
		\bigg[
		U_0^{}V_0^T+\frac{1}{2}U_0(\mathcal{F}^T-\mathcal{F}^{})V^T_0
		\bigg]_{\mu_a \mu_b},
	\end{aligned}
\end{equation}
and for the case of two-matter fermion excitations, i.e., 
$ |M_{\mu_1}^{xy}\rangle=\hat{\overline{a}}^{\dagger}_r
\hat{\overline{a}}^{\dagger}_s|\mathcal{M}_{\mu_1}^{xy}\rangle $,
\begin{equation}
	\label{twoMatterExci}
	\begin{aligned}
		&\langle P_{\mu_1}^{xy}|\hat{\sigma}_{a}^z|G\rangle=
		-i\sqrt{|\det(\mathcal{X})|}\bigg[
		\mathcal{Y}\mathcal{X}^{-1}\bigg]_{rs},\\
		&\langle P_{\mu_1}^{xy}|\hat{\sigma}_{b}^z|G\rangle=
		-i\sqrt{|\det(\mathcal{X})|}\ 
		\bigg[
		(U_0\mathcal{X}^{-1})_{\mu_a,r}(V_0\mathcal{X}^{-1})_{\mu_b,s}
		-(U_0\mathcal{X}^{-1})_{\mu_a,s}(V_0\mathcal{X}^{-1})_{\mu_b,r}
		+(\mathcal{Y}\mathcal{X}^{-1})_{rs}(U_0^{}V_0^T+\frac{1}{2}U_0(\mathcal{F}^T-\mathcal{F}^{})V^T_0)_{\mu_a\mu_b}
		\bigg],
	\end{aligned}
\end{equation}
where the last matrix element corresponds to the sum of 40 terms, akin to the expression in  Eq.(\ref{three_matter_mat_element}). In equations (\ref{ZeroMatter}) and (\ref{twoMatterExci}), the matrices $\mathcal{X}$ and $\mathcal{Y}$ are defined as 
\begin{equation}
	\label{bogolibov}
	U=\begin{pmatrix}
			\mathcal{X}^*&\mathcal{Y}^*\\
			\mathcal{Y}&\mathcal{X}
		\end{pmatrix},
\end{equation}
where $U$ represents a unitary transformation that diagonalizes  Hamiltonian (\ref{Ham_compact_mag_complex})\cite{Knolle_Chalker_Moessner2015}.

In the numerical computation of the non-linear responses of the Kekulé-Kitaev model, we consider the gauge configuration with the topological label $(+1, -1)$ for the ground state, see Fig.~\ref{topological_loop}. In the presence of a uniform magnetic field, we choose the standard gauge configuration for the ground state. While in the presence of a staggered magnetic field, the 0-flux state labeled $(+1, +1)$ is no longer belongs to the ground state manifold due to the increase in energy\cite{Negahdari_Langari_2023}. Consequently, we consider the gauge configuration  labeled by $(+1,-1)$ for the ground state.
 
Utilizing the translational symmetry of the ground state simplifies the calculation of non-linear response. For instance, the quantity  $  \Upsilon^{(1),z}_{zzz}(\tau_2,\tau_1,0) $ can be expressed as
\begin{align}
	\label{phy_equ_R12}
	\Upsilon^{(1),z}_{zzz}(\tau_2,\tau_1,0)=&3N\sum_{\mu_i}
	e^{i[E({1_1\mu_i})-E_2]\tau_2}\bigg\{
	\sum_{p=1}^{3N} e^{i\varepsilon_p^{(1_1)})\tau_1}
	\langle G|\hat{Z}_{1_1}
	|{P_{1_1}}\rangle
	\langle{P_{1_1}}|\hat{Z}_{\mu_i}
	|{Q_{1_1\mu_i}}\rangle
	\sum_{r=1}^{3N}
	e^{-i\varepsilon_r^{(1_1)}(\tau_1+\tau_2)}
	\langle{Q_{1_1\mu_i}}|\hat{Z}_{\mu_i}
	|{R_{1_1}}\rangle
	\langle{R_{1_1}}|\hat{Z}_{1_1}
	|G\rangle\nonumber\\
	&+
	\sum_{p=1}^{3N} e^{i\varepsilon_r^{(\mu_i)}\tau_1}
	\langle G|\hat{Z}_{\mu_i}
	|{P_{\mu_i}}\rangle
	\langle{P_{\mu_i}}|\hat{Z}_{1_1}
	|{Q_{1_1\mu_i}}\rangle
	\sum_{r=1}^{3N}
	e^{-i\varepsilon_r^{(1_1)}(\tau_1+\tau_2)}
	\langle{Q_{1_1\mu_i}}|\hat{Z}_{\mu_i}
	|{R_{1_1}}\rangle
	\langle{R_{1_1}}|\hat{Z}_{1_1}
	|G \rangle\bigg\},
\end{align}
where $\mu_i\neq1_1$.
As a reminder, $\mu$  represents the unit cell label, while $\mu_i$  indicates $z$-links, see Fig.~\ref{Majorana_paring}.

 \begin{figure}[!t]
	\includegraphics[width=0.75\columnwidth]{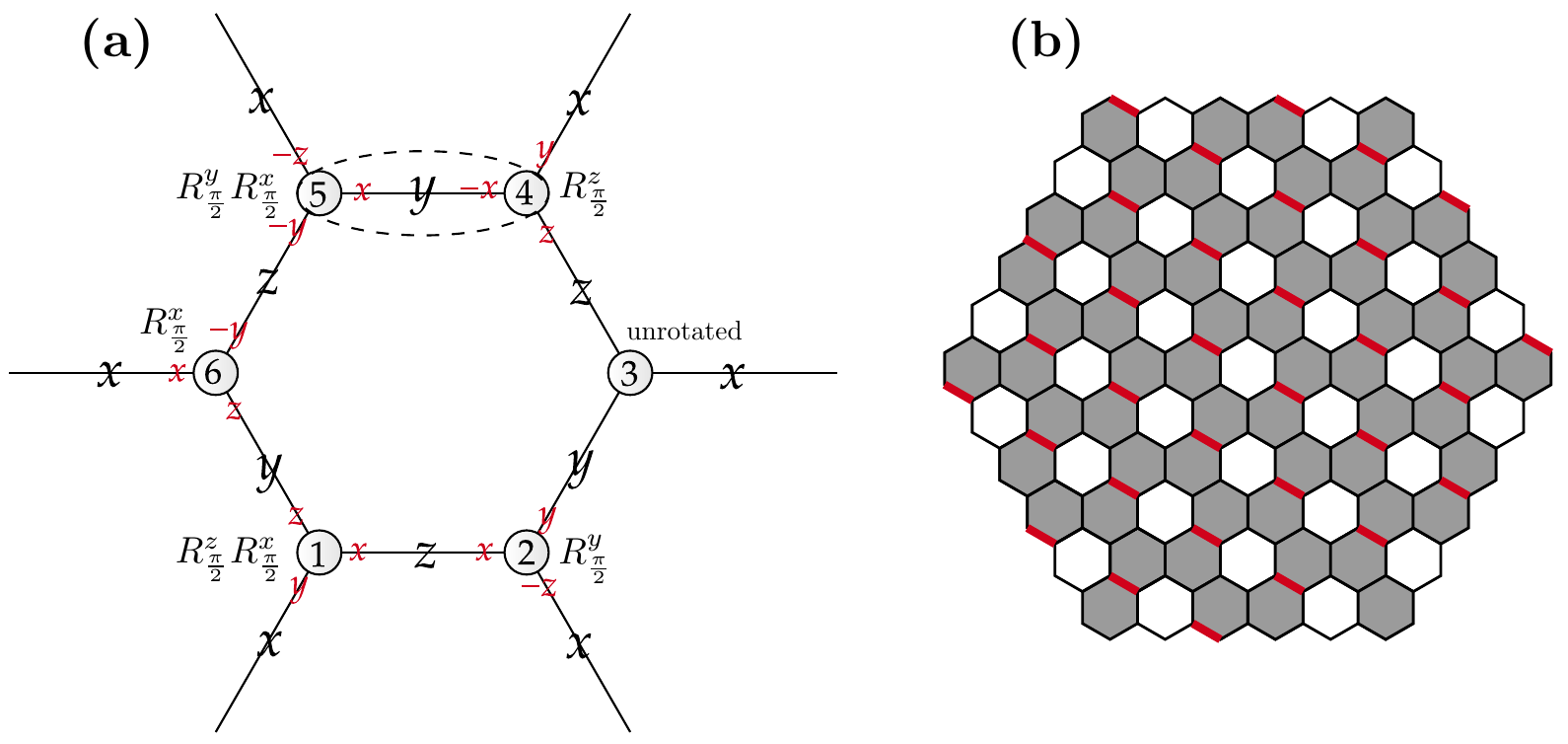}
	\caption{(a) The unit cell of the Kekulé-Kitaev model. The global spin transformation that maps the Kekulé-Kitaev model to the Kitaev model consists of on-site spin rotations with $R_{\phi}^{\alpha}$ operators specified near each site of the unit cell. Two rotations are applied at sites 1 and 5. The black component written at the center of each bond indicates the type of Ising interaction in Kekulé-Kitaev model, while the red components near each site represent the obtained components of the spin operators after rotation. Under this transformation, the sign of  interaction on the bond marked by the dashed oval is reversed, which can be absorbed into the gauge field on that bond. Through this mapping, the ground state of the Kekulé-Kitaev model transforms to an excited state of the Kitaev model with the flux configuration shown in (b). The blak and red bonds indicate  $u_{jk}=+1 \; \mbox{and}\; u_{jk}=-1$, respectively.}
	\label{KekuleToKitecvModel}
\end{figure}

\section{Mapping the spectrum of Kekulé-Kitaev model to the Kitaev model}\label{mapping}
 At the isotropic coupling $J_x=J_y=J_z=J$, the unitary spin rotations on each site of the unit cell, as illustrated in Fig.~\ref{KekuleToKitecvModel}(a), map the Kekulé-Kitaev model to the Kitaev model, with a sign change of the upper bond of the unit cell.
In  Fig.~\ref{KekuleToKitecvModel}(a), the operators
$ R_{\phi}^{\alpha}=e^{-i\hat{\sigma}^{\alpha}\phi/2}$, written near each site of the unit cell, indicate a spin rotation around the axis 
$\alpha $ by an angle $\phi $. The spins are rotated as $R_{\phi}^{\alpha}\hat{\sigma}^{\beta}_jR_{\phi}^{\alpha\dagger}$.

In the Kitaev model, bonds with the interaction type $J\hat{\sigma}^{\alpha}_j\hat{\sigma}^{\alpha}_k$ are aligned parallel across the entire lattice. Accordingly, in this transformation,  we take the direction of interactions at site 3 as the reference direction. Thus,  site 3 is unchanged during the transformation. However, under this transformation, the sign of interaction on the bond indicated by the dashed oval (upper bond of unit cell) in Fig.~\ref{KekuleToKitecvModel}(a) is reversed. Prior to the transformation, the interaction of this bond is
$J\hat{\sigma}_5^y\hat{\sigma}_4^y$, while it becomes $ -J\hat{\sigma}_5^x\hat{\sigma}_4^x$ after the transformation. Given the gauge structure of  the system, this additional sign can be absorbed into the gauge fields.
Therefore, if we choose an arbitrary gauge configuration
\begin{equation}
\{u_{12},u_{32},u_{34},u_{54},u_{56},u_{16}\},
\end{equation}
for the unit cell in the Kekulé-Kitaev model, after applying this transformation, we adopt the gauge configuration
\begin{equation}
	\{u_{12},u_{32},u_{34},-u_{54},u_{56},u_{16}\},
\end{equation}
for the Kitaev model. In particular, this implies that the ground state of the Kekulé-Kitaev model with the standard gauge configuration
$\{u_{jk}=+1,\ \forall j\in{\text{odd}},k\in{\text{even}}\}$ is transformed into an excited state of the Kitaev model with the flux configuration shown in Fig.~\ref{KekuleToKitecvModel}(b), characterized by the flux density  $\rho_f=2N/3N=2/3$. Notably, the ground state of the Kekulé-Kitaev model, as illustrated in Fig.~\ref{energy_bands}(a), has six energy bands in momentum space, and the flux configuration shown in Fig.~\ref{KekuleToKitecvModel}(b) similarly results in six energy bands for the Kitaev model.

\end{widetext}
%

\end{document}